\documentclass[twocolumn,iop,revtex4]{openjournal}

\usepackage{newtxtext,newtxmath}

\usepackage[T1]{fontenc}

\DeclareRobustCommand{\VAN}[3]{#2}
\let\VANthebibliography\thebibliography
\def\thebibliography{\DeclareRobustCommand{\VAN}[3]{##3}\VANthebibliography}

\usepackage{graphicx}	
\usepackage{amsmath}	
\usepackage{soul}

\usepackage{natbib} 
\usepackage{amssymb}
\usepackage[title]{appendix}
\usepackage{hyperref}	
\hypersetup{colorlinks=true,linkcolor=blue,citecolor=blue,filecolor=blue,urlcolor=blue}

\newcommand{\Msun}{M_{\odot}}

\begin{document}

\title[Halo Masses]{Halo Properties from Observable Measures of Environment: I. Halo and Subhalo Masses}

\author{
Haley Bowden,$^{1}$\thanks{E-mail: hbowden@arizona.edu}
Peter Behroozi,$^{1,2}$
Andrew Hearin,$^{3}$}

\affiliation{
$^{1}$Department of Astronomy and Steward Observatory, University of Arizona, Tucson, AZ 85721, USA
\\
$^{2}$Division of Science, National Astronomical Observatory of Japan, 2-21-1 Osawa, Mitaka, Tokyo 181-8588, Japan
\\
$^{3}$High-Energy Physics Division, Argonne National Laboratory, Argonne, IL 60439, USA
}

\begin{abstract}
The stellar mass – halo mass relation provides a strong basis for connecting galaxies to their host dark matter halos in both simulations and observations. Other observable information, such as the density of the local environment, can place further constraints on a given halo’s properties. In this paper, we test how the peak masses of dark matter halos and subhalos correlate with observationally-accessible environment measures, using a neural network to extract as much information from the environment as possible. For high mass halos (peak mass $>10^{12.5} \Msun$), the information on halo mass contained in stellar mass--selected galaxy samples is confined to the $\sim$ 1 Mpc region surrounding the halo center. Below this mass threshold, nearly the entirety of the information on halo mass is contained in the galaxy's own stellar mass instead of the neighboring galaxy distribution. The overall root-mean-squared error of the best-performing network was 0.20 dex. When applied to only the central halos within the test data, the same network had an error of 0.17 dex. Our findings suggest that, for the purposes of halo mass inference, both distances to the $k$th nearest neighbor and counts in cells of neighbors in a fixed aperture are similarly effective measurements of the local environment.
\end{abstract}




\section{Introduction}\label{sec:Intro}
In the Lambda Cold Dark Matter ($\Lambda$CDM) paradigm, all galaxies form at the centers of dark matter halos. Structure formation within this framework proceeds hierarchically, such that small halos merge into larger ones, becoming satellites of the larger central halo. A galaxy's formation and evolution are tied to the central or satellite halo in which it lives (see \citealt{Wechsler_2018} for a review). This connection has fueled interest in learning about halos to better understand the physics of galaxy formation and conversely to use observations of galaxies to constrain the structure and abundance of dark matter. The galaxy-halo connection persists across a wide range of scales from small halos hosting dwarf galaxies ($\lesssim 10^{11} \Msun$) to massive halos hosting galaxy clusters ($\gtrsim 10^{14} \Msun$). Across these mass scales, accurate measurements of halo masses have many potential applications.

At the low-mass end, host halo masses are vital for understanding the physical processes that govern the formation and evolution of galaxies. The halo mass is directly linked to the amount of baryonic matter available for star formation, and it is also a primary driver of the growth and feedback processes that shape the galaxy \citep[e.g.,][]{UM,Moster2018}. In particular, there is a well-studied relationship between stellar mass and host halo mass, known as the stellar mass – halo mass relation (SHMR; see e.g., \citealt{Wechsler_2018} and references therein).

The exact mapping between stellar mass and halo mass, as well as the amount of scatter in this relationship, tells us about the efficiency of the gas-to-stars conversion processes (\citealt{Moster2020,Munshi2021}). Hence, investigations into the masses of halos hosting dwarf galaxies are typically aimed at determining: 1) the halo mass at which galaxies start forming, and/or 2) the scatter in the SHMR for dwarf galaxies. Overall, more widespread access to halo masses will lead to an improved understanding of the physics involved and the creation of galaxy models that more accurately correspond to the real Universe.

At slightly higher halo masses, accurate measurements are useful in interpreting observations. For example, in studies aiming to probe the contents or extent of the circumgalactic medium (e.g., \citealt{Zhang2020,Werk2014}), knowledge of the halo mass is important for understanding the size of the virialized region around the galaxy. The gas surrounding the target galaxy is often probed by observing a background quasar or galaxy (see \citealt{Tumlinson2017} for a review). In these cases, accurate measurement of a foreground halo's mass is essential for estimating the halo radius, and thus determining whether a given sight line is probing the circumgalactic medium inside the dark matter halo or the intergalactic medium outside the halo \citep{Peeples2019}. In this way, the dark matter halo mass gives us a way of interpreting observations that would otherwise be ambiguous. For studies like these that use the halo mass to understand galaxy formation, $\sim0.3$ dex uncertainties on observed stellar masses (see e.g., \citealt{Conroy_2013}) mean that measuring halo masses more accurately than $\sim$ 0.2 dex typically does not lead to significantly tighter constraints on galaxy formation. 

Halo masses, particularly on the scale of galaxy groups and clusters, also carry information about cosmological parameters and large-scale structure \citep{Weinberg2013}. Hence, there is widespread interest in measuring halo masses for clusters to determine the total number of halos above a given mass and thereby constrain the matter density in the universe, as well as the normalization of the power spectrum (\citealt{Ho_2019,PLANCK2018}). Cosmology applications typically require as accurate constraints as possible on halo masses to achieve the best constraints on cosmological parameters.

Despite their relevance across several fields, obtaining accurate host-halo masses remains a challenge across the entire mass spectrum. Additionally, as there are distinct goals relevant to the different mass regimes, many previous techniques have been optimized for specific subsets of halo mass. 

Over the past two decades, numerous studies have connected galaxy properties with host halo mass using empirical modeling techniques (e.g., halo occupation distributions and empirical modeling). These techniques use abundance and clustering data to match galaxies to halos and subhalos (see \citealt{Wechsler_2018} and references therein). Halo abundance matching can be highly effective \citep{Old2014}, but is limited by the intrinsic scatter between the matched galaxy property and halo mass. More recent efforts (e.g., \citealt{UM,Moster2018}) have sought to eliminate this issue by empirically modeling the evolving connection between galaxies and halos over cosmic time. This is the only existing technique that can be applied to satellite halos as well as central halos.

For massive halos, particularly on galaxy cluster scales, there are three major alternative approaches for measuring individual halo masses. One approach relies on satellite kinematics from spectroscopic surveys (e.g., \citealt{Bosch2008,More2011,Tinker_2021}), where satellite positions and velocities are used to determine which galaxies are satellites and which are unbound, thereby determining the extent of the halo. This technique tends to work down to halo masses of $\sim 10^{12} \Msun$, below which there are too few observable satellite galaxies per halo in extragalactic surveys. Similarly, spectroscopic, as well as photometric data, has been used to identify galaxy groups (e.g., \citep{Wang_2020}, while tools such as redMaPPer \citep{Rykoff2014} rely on photometry alone to identify clusters via satellite number counts (i.e., ‘richness’). \citet{Old2014} and \citet{Old2015} provide an overview of these approaches, finding that dynamical and galaxy-based techniques can provide accurate measurements of halo mass to a factor of $\sim 2$ in the $M > 10^{14} \Msun$ regime, with significantly larger errors at lower halo masses.  The third technique estimates halo masses using evidence of a hot halo through X-ray measurements (\citealt{Mantz2016,Giles2017}) or the Sunyaev-Zeldovich effect (\citealt{Sunyaev1972,PLANCK2018}). Like the satellite-based approaches, this technique is less effective at lower halo masses.

Machine learning (ML) techniques provide a way to circumvent the limitations of traditional techniques and incorporate high-dimensional data to develop models of a wide variety of physical phenomena. Over the last decade, ML methods have been used to extract information from observations (or simulated observations) to enhance halo mass estimates. On the massive cluster scale, several studies have used ML to measure halo masses using dynamical and/or X-ray data (\citealt{Ntampaka2015,Ntampaka2016,Ntampaka2019,Armitage2019,Ho_2019}). Other ML studies have incorporated different types of observables, such as photometric, structural, and kinematic data of the hosted galaxy \citep{VonMarttens2021} or a diverse set of galaxy and group features \citep{Calderon2019}, finding an improvement in accuracy over traditional halo abundance matching and dynamical mass estimates when applied to simulated datasets. 

Previous local-environment-based methods have tended to consider information that is difficult to retrieve fully from observations, with, for example, \citet{Calderon2019} using all nearby galaxies without imposing a stellar mass limit or \citet{Villanueva_Domingo_2022} using 3D distances between galaxies. In this paper, we limit our data to realistically observable information by restricting the stellar mass range of our sample as well as the available galaxy properties and position information. 

Our major sources of information on halo properties are based on two standard environmental measures in observational work. Simulations suggest that the local environment contains information about halo properties (\citealt{Lee2017,Behroozi2022}). However, there is no standard environmental indicator, as some have been found to have advantages and disadvantages for different research goals and different sets of observational data. This paper focuses on two popular methods for probing the density of galaxies: 1) the distance to the $k$th nearest neighbor ($k$NN), and 2) counts of neighbors within a fixed aperture (see \citealt{Muldrew2012} for a review of both techniques). We compare the effectiveness of the two separate probes over different halo mass regimes to better understand the environmental information provided by each.

Many popular environmental measures (e.g., the two-point correlation function) are functions of the distances to the $k$ nearest neighbors. These distances are usually defined as the projected distances (i.e., the 2D comoving separations) to neighboring galaxies within a redshift separation of typically $ \lesssim$ 1000 km s$^{-1}$. Neighbor distances have been used by a number of studies (e.g., \citealt{Muldrew2012,Li2011,Cowan2008}). \citet{Muldrew2012} explored the effectiveness of different nearest neighbors-based statistics and found nearest neighbors to be an effective probe of the local environment, with fixed aperture methods being more effective at measuring the large-scale environment. However, previous studies have focused on small values of $k$ ($<10$), while more distant neighbors still potentially contain information about the target halo.  With ML techniques, it is simple to retain a substantial number of neighbors and search for an approximate optimal mass estimator over the large resultant parameter space. Yet, even when retaining a large number of neighbors, this probe might break down on cluster mass scales, where we expect many satellites. In these cases, the nearest neighbors probe could inadequately probe the full extent of the halo (e.g., if the number of satellites is $\gg$ 50) or fail to separate close satellites from other neighbors given limited redshift information.

The second method probes a fixed length scale for all galaxies, rather than probing a length scale dependent on the density of the environment. This is usually done by defining a cylinder around a target object with fixed projected distance ($\sim$ 0.5 – 5 $h^{-1}$Mpc) apertures (e.g., \citealt{Grutzbauch2011}) or annuli \citep{Wilman2010}, within a certain redshift offset (ranging from 500 to 6000 km s$^{-1}$). This method is based on the correlation between richness and halo mass (e.g., \citealt{Yee2003} and \citealt{Old2014}), which suggests that counts of galaxies in cylinders should scale with halo mass, particularly at cluster mass scales. Large-scale bias is also expected to scale with halo mass at high masses (see \citealt{Wechsler_2018}). Given these scaling relationships, we expect that cylinder counts would perform best for high-mass halos ($M > 10^{13} \Msun$).

The goals of this paper series are to extract the relevant information from these environmental measures to provide estimates of halo properties (including mass, concentration, and assembly history), analyze the correlations between these properties and galaxy properties, and determine the observational metrics that are most sensitive to a given halo property. The focus of this paper is on halo and subhalo masses and is organized as follows. In Section \ref{sec:Data}, we discuss the simulated halo and galaxy properties used. Section \ref{sec:Methods} gives an overview of the sources of environmental information (Section \ref{subsec:Environment}), sample statistics (Section \ref{subsec:Sample}), and the methods by which we develop and train a neural network (Sections \ref{subsec:Preprocess} and \ref{subsec:Network}). We evaluate the performance of the trained networks in Section \ref{sec:Results}. In Section \ref{sec:Discussion}, we summarize our results and discuss future applications of this technique. Throughout, we adopt a standard $\Lambda$CDM cosmology with $(h, \Omega_m, \sigma_8, n_s) = (0.678, 0.307, 0.823, 0.96)$.

\section{Data}\label{sec:Data}

\subsection{Overview}\label{subsec:DataOver}

To estimate halo mass, we used galaxy properties that are both observable and confidently simulated, including 1) projected distances to the target galaxy's neighbors within bins in redshift space, and 2) cumulative number densities of the target galaxy's stellar mass and the stellar masses of its neighbors. Simulated halo properties are from the \textit{Small MultiDark Planck (SMDPL)} and \textit{Bolshoi-Planck} cosmological simulations (Section \ref{subsec:HaloProps}; \citealt{Bolshoi,Rodr_guez_Puebla_2016}). Individual galaxy properties were assigned to halos using the \textsc{UniverseMachine} empirical model (Section \ref{subsec:GalProps}; \citealt{UM}).

\subsection{Halo Properties}\label{subsec:HaloProps}

The neural network was trained on $z=0$ halo properties extracted from the SMDPL simulation \citep{SMDPL}, which has a periodic (400 $h^{-1}$Mpc)$^3$ volume and 3840$^3$ particles, corresponding to a mass resolution of $9.63 \times 10^7 h^{-1} \Msun$ per particle and a force resolution of 1.5 $h^{-1}$kpc. This simulation adopts a flat $\Lambda$CDM cosmology with $(h, \Omega_m, \sigma_8, n_s) = (0.678, 0.307, 0.823, 0.96)$, consistent with the most recent Planck results \citep{PLANCK2018}. We assume the same cosmology throughout this work. 

Halo finding was conducted using \textsc{Rockstar} \citep{Rockstar} and merger trees were constructed with the \textsc{ConsistentTrees} code \citep{ConsistentTrees}. Halo masses were defined using the \citet{Bryan1998} virial spherical overdensity criterion ($\rho_\mathrm{vir}$). Throughout, we consider peak halo mass ($M_p$), defined as the maximum mass of the halo across all prior snapshots, rather than the current halo mass (at the time of the snapshot) as it is more closely linked to stellar mass \citep[see][for a review]{Wechsler_2018}. 

It is essential to test the neural network's performance on distinct data from the training sample. For this purpose, we used the smaller \textit{Bolshoi-Planck} dark matter simulation box \citep{Bolshoi}, which has a periodic (250 $h^{-1}$Mpc)$^3$ co-moving volume with 2048$^3$ particles, corresponding to a mass resolution of $1.55 \times 10^8 h^{-1} \Msun$ per particle and a force resolution of 1.0 $h^{-1}$kpc. The simulation uses a similar cosmology to the SMDPL simulation with $(h, \Omega_m, \sigma_8, n_s) = (0.68, 0.30711, 0.82, 0.96)$. Given its relatively small size, the \textit{Bolshoi-Planck} box contains a limited sample of high-mass halos (fewer than 1000 with peak halo mass $M_p > 10^{14} \Msun$). By using the SMDPL simulation as the training sample, we ensured that the network has a sufficient number of high-mass halos on which to train (more than 3000 with $M_p > 10^{14} \Msun$). 

 \begin{figure}
     \centering
     \includegraphics[width=0.45\textwidth]{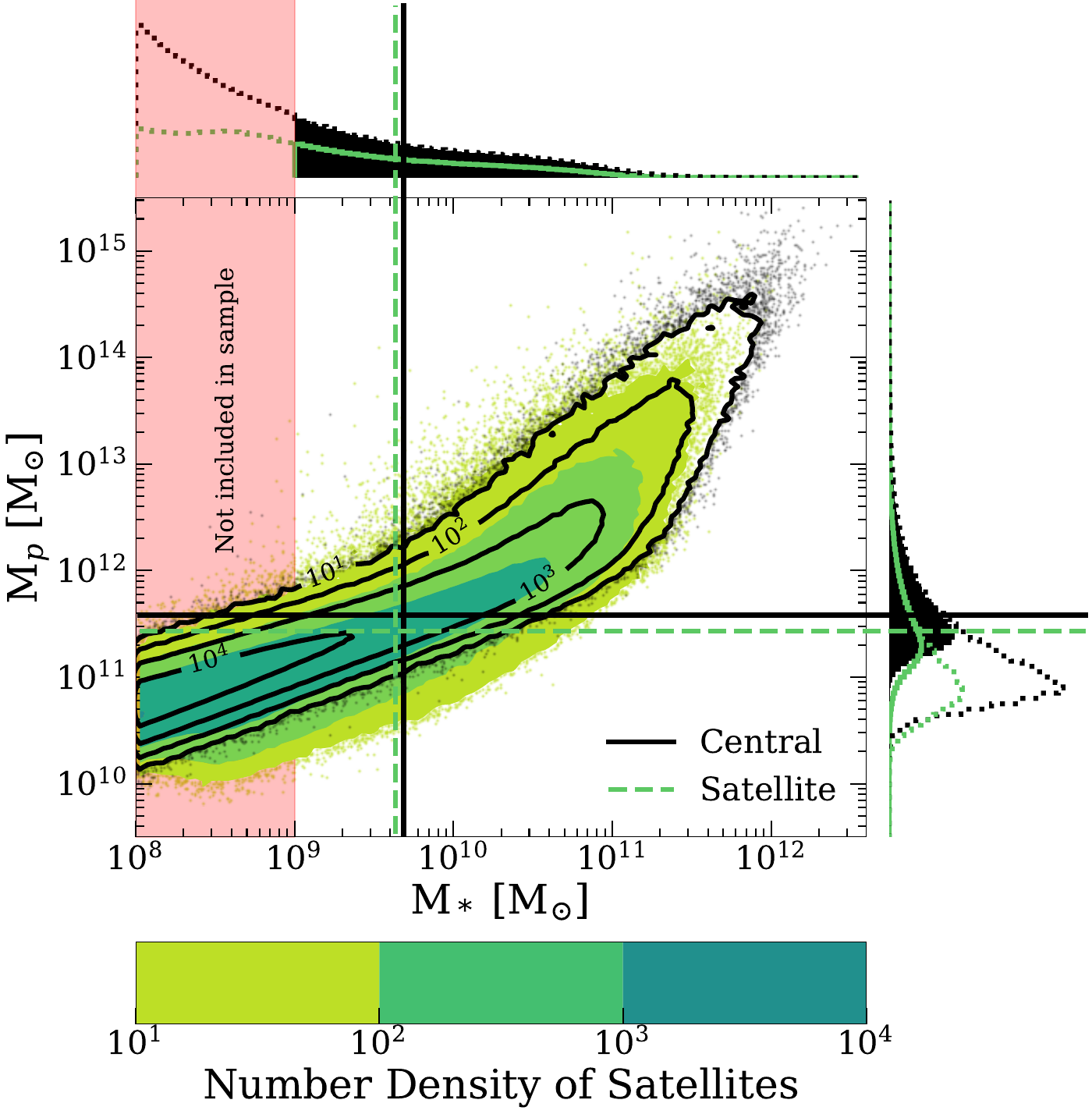}
     \caption{The distribution of \textsc{UniverseMachine} galaxies in the SMDPL simulation box with respect to observed stellar mass and the peak mass of their host dark matter halo. The filled green contours show the distribution of the satellite population, while the black contour lines show the central population. The black (green) histograms show the central (satellite) stellar and halo mass functions individually, with the black solid lines denoting the median of each distribution.The majority of galaxies and host halos fall in the lower mass regions, but each also exhibits a tail out to high masses. Note that limiting the galaxy sample to those with stellar mass $M_* > 10^9 \Msun$ selects the upper portion of the stellar mass distribution but still covers a wide array of halo masses. The red highlighted region and the dotted histogram outlines show the population distributions if galaxies with stellar masses down to $M_* = 10^8 \Msun$ were included.)
     }
     \label{fig:MassDist}
 \end{figure}

\subsection{Galaxy Properties}\label{subsec:GalProps}

The dark matter halos were populated with galaxies using \textsc{UniverseMachine} \citep{UM}. \textsc{UniverseMachine} is an empirical model that uses a Markov Chain Monte Carlo algorithm to constrain how galaxy star formation rates depend on halo mass, halo growth rates, and cosmic time. The algorithm constrains the galaxy--halo relationship by requiring the overall population to match observations of: 1) the stellar mass function ($z\sim 0-4$), 2) cosmic star formation rates ($z\sim 0-10$), 3) specific star formation rates ($z\sim 0-8$), 4) UV luminosity functions ($z\sim 4-10$), 5) quenched fractions ($z\sim 0-4$), 6) median UV-stellar mass relations ($z\sim4-10$), 7) correlation functions for quenched and star-forming galaxies ($z\sim 0-1$), and 8) the dependence of the quenched fraction on environment ($z\sim 0$). Appendix C in \cite{UM} contains the full references for these observational constraints.

We selected one snapshot at $z=0$ from the \textsc{UniverseMachine} mock galaxy-catalog for each of the \textit{Bolshoi-Planck} and SMDPL boxes. From these snapshots, we extracted galaxy positions, velocities, and observed stellar masses. \textsc{UniverseMachine} models both true and observed stellar masses. The observed stellar mass values were adjusted from true stellar masses, taking into account systematic offsets between true and observed stellar masses as well as the random scatter in observed stellar masses \citep{UM}. 

Although we exclusively used \textsc{UniverseMachine} stellar masses in this paper, the method is designed for eventual application to stellar masses derived from observations. Hence, observed galaxy stellar masses from the \textsc{UniverseMachine} were converted to cumulative number densities as measurements of observed stellar masses are model and calibration dependent. Stellar masses are primarily derived from light using spectral energy distribution fitting, which depends on assumptions such as the star formation history of the galaxy and the relevant dust attenuation law. Different model assumptions produce inconsistent stellar masses \citep[e.g.][]{Conroy2013,Madau2014,Mobasher2015}. 

To translate stellar masses to cumulative number densities we first rank order the galaxies within the simulation box, with the most massive being assigned a rank of one. Then we normalized these rankings by the box volume. This process removes the largest systematic offsets in stellar mass between different models, and allows a neural network trained on one dataset to be more robust when applied in the future to data. 

While \textsc{UniverseMachine} provides star formation rates, the relationship between star formation rates and halo properties is not as robustly established as that for stellar masses, and so we take a conservative approach by not using star formation rates as inputs here.

\section{Methods}\label{sec:Methods}

In Section \ref{subsec:Environment}, we discuss how the local galaxy environment is defined. Section \ref{subsec:Sample} covers the statistics of the galaxy and halo populations considered. We then preprocess the data in Section \ref{subsec:Preprocess}. Section \ref{subsec:Network} describes the general architecture of the neural networks and the training process. We define our galaxy sample as all galaxies within the simulation box with $M_* > 10^9\Msun$, corresponding typically to halos with $M_p \gtrsim 10^{11.5} \Msun$. This includes both central and satellite galaxies, as the two categories cannot be perfectly separated using observations.

\subsection{Sources of Environmental Information}\label{subsec:Environment}

The input layer of each network was composed of the stellar mass of the target object concatenated with a data vector of environmental information. The following sections describe how these data vectors were defined. 

\subsubsection{Distances to Nearest Neighbors}\label{subsubsec:Environment1}

We searched for the fifty closest neighboring galaxies with a redshift offset of $<1000$ km s$^{-1}$ and did not consider galaxies outside this cut as potential neighbors. This cut eliminated neighbors with a high redshift separation from the target galaxy without excluding the majority of a galaxy's potential satellites from consideration. A 1000 km s$^{-1}$ cut corresponds to the virial velocity of clusters with $M_p\sim 10^{14} \Msun$. Some galaxies within a cluster may be excluded with this cut, but including galaxies at a larger velocity separation has the potential to introduce more noise from projection effects for neighboring galaxy distances at the low-mass end.

We searched for the nearest neighbors to each galaxy within these redshift cuts, where the projected distance to a neighbor is measured in the $x-y$ plane. We imposed a stellar mass cut on neighbors such that a neighbor must have a stellar mass no less than 1) 1.5 dex below that of the galaxy under consideration, or 2) $10^9 \Msun$, whichever is highest. For each galaxy, we considered the fifty nearest neighbors that met those criteria, covering projected distances up to $\sim$11$h^{-1}$Mpc in the case of the most isolated galaxies. This information was given to the neural network as a vector of stellar masses and projected separations from the target. The stellar mass of each neighboring galaxy is scaled according to the number density of objects as described in Section \ref{subsec:GalProps}. We refer to the projected separations between target and neighbor as the distance to the $k$th nearest neighbor, where $k$ is the rank of the neighbor in separation from the target.

In addition, we considered the redshift separations between the target and its neighbors as potential inputs. However, no significant changes in network performance were noted with these additional inputs, and the additional inputs resulted in increased network training times (see Appendix \ref{sec:AppB}). Thus, we excluded this data from all further analyses.

\subsubsection{Counts in Cylinders}\label{subsubsec:CoC}

Counts in cylinders are an environmental measure with a fixed spatial scale, unlike the $k$th nearest neighbors measure, which covers an area dependent on the local density of galaxies. This method is also distinct from the nearest neighbors method described in the previous section as 1) it does not consider the stellar mass of the neighbors beyond whether they meet the minimum mass requirement and 2) it makes use of redshift information.

To measure counts in cylinders, we selected circular apertures with radii of 0.5 $h^{-1}$Mpc, 1 $h^{-1}$Mpc, 2 $h^{-1}$Mpc, and 5 $h^{-1}$Mpc. These values are spaced in $\sim 0.3$ dex intervals in radius, corresponding to changes in virial radius associated with $\sim 1$ dex intervals in halo mass. These aperture sizes were selected to provide sensitivity to a wide range of halo masses.

We retained the same stellar mass cuts as in the $k$th nearest neighbors case. A search was then performed to find the number of neighboring galaxies within each cylinder. Once this was complete, the neighbors were further split into bins by absolute redshift separation from the target. Bin widths are $|\Delta z|$ = 250 km s$^{-1}$ each and together cover separations of up to $|\Delta z| = 2000$ km s$^{-1}$. Splitting the data into narrower bins provides information to better exclude sources at larger redshift separations that have low projected distances from the target galaxy. The bin spacing of 250 km s$^{-1}$ was chosen to retain this redshift separation information while also reducing the Poisson noise that would result from using narrower bins and avoiding the additional complexity of a neural network with significantly more velocity bins. Appendix \ref{sec:AppB} describes a version of this method with no redshift binning for more direct comparison to the nearest neighbors method.

\subsection{Sample Statistics}\label{subsec:Sample}

We selected 2,877,669 objects from the SMDPL simulation box that met our sample criteria, including both central and satellite halos. The distribution of the peak halo masses and $z=0$ stellar masses of these galaxies are shown in Figure \ref{fig:MassDist}. The galaxies in the SMDPL box make up the training and validation data sets for the neural networks.

The \textit{Bolshoi-Plank} simulation box contains similar halo and galaxy mass distributions to SMDPL. From this box, we selected 695,554 galaxies meeting the stellar mass cutoff criterion, which is approximately 25\% of the size of the SMDPL dataset. The objects in the \textit{Bolshoi-Planck} box make up the test data set.
 
\subsection{Pre-processing}\label{subsec:Preprocess}

Before neural network training, we normalized and scaled the inputs (commonly known as features) and outputs (labels) of the network. This is essential as neural networks can be sensitive to the scale of data, and having input and output data covering several orders of magnitude in scale can result in poor model performance \citep{CAMES}. Each property was individually standardized to have a mean of zero and a standard deviation of one. For stellar and halo masses, this was performed on a base-ten logarithmic scale, while the distances to the fifty nearest neighbors and the counts in cylinders were scaled linearly. 

In addition, we wanted the network to prioritize halos at the extremes of halo mass. The vast majority (98\%) of SMDPL halos had peak masses between $10^{10.5} M_\odot$ and $10^{13} M_\odot$. Less than 0.05\% of halos fell below this mass range. 1.9\% of galaxies had halo masses above this range and only 0.12\% had halo masses of $10^{14} M_\odot$ or above. Due to this bias in halo number density when separated by mass bin, the default fitting procedure prioritizes typical-mass halos. To counteract this effect, and incentivize the network to also fit the halos in the less populated mass bins, we tested weighting the data conditionally by halo mass. 

To weight the data, we first separated the training and validation data into 25 bins by halo mass. We then found the number of objects within these bins. These count values were assigned to the median halo mass in the bin and a linear interpolation from these counts was used to calculate an approximate normalized halo mass function where $n(M_{halo})$ is the number density of halos at a given halo mass in the simulation. The weight ($W$) assigned to a given halo is defined according to: 
\begin{equation}
    W(M_{halo}) = \frac{1}{\sqrt{n(M_{halo})}}.
    \label{Weights}
\end{equation}
In the following sections, we consider a network trained on an unweighted dataset and one trained on a sample weighted as in Eq.\ \ref{Weights} to prioritize objects with extreme halo masses.

\subsection{The Neural Network}\label{subsec:Network}

Using a neural network, we created an approximate mapping between the inputs (observable data about a galaxy) and outputs (the peak mass of the host halo). We used supervised learning to train the network, i.e., the network is iteratively trained to minimize the error of output predictions by using the provided input-output pairs to adjust its internal weights.  The trained network then acts as a function that takes data from outside the training sample as input and can make new predictions based on that data.

We developed networks based on three major input types consisting of a fixed length vector of 1) the distances and stellar masses of the target galaxy’s fifty nearest neighbors, 2) counts of galaxies in cylindrical apertures around the target galaxy, or 3) the combination of the previous inputs. Each network also takes the stellar mass of the target galaxy as an additional input. The three networks are all designed to predict the halo mass of the target galaxy’s host halo. 

We used \textit{Keras} \citep{keras}, a popular open-source machine learning library, with a \textit{Tensorflow} backend \citep{tensorflow}, to construct and train our neural network. Each network uses a fully-connected network structure, which means every node in a layer receives input from all the nodes in the previous layer. A non-linear activation function is applied to each node after weighting. The nonlinearity of the network and the large number of flexible weights allow it to learn complex relationships between the input data and the target outputs.

For training, the SMDPL box was split in two along the x-axis, with 70\% of the volume (2,048,724 galaxies) making up the training data set and the remaining 30\% (828,945 galaxies) reserved for model validation. The network was provided with fully-labeled data (i.e., including true values for outputs in addition to inputs) during the training stage. The validation data is used to evaluate the performance of the network during the training process to prevent it from overfitting the training set (see \citealt{Xu2018}). The test set, composed of galaxies from the \textit{Bolshoi-Planck} box, was not viewed by the network during training. Instead, this data was set aside to evaluate the performance of the final trained networks on data not seen before.

In addition to the numerous trainable parameters, there are several non-trainable parameters that describe the model architecture, design, and training. We ran a search comparing model performances with different sets of architectures and hyperparameters. The parameters considered are described in the remainder of this section and summarized in Table \ref{tab:Fixed}. Model parameters were chosen primarily based on their accuracy on the validation dataset. When different models obtained similar performances, we favored the models with fewer trainable weights and shorter training times. The values chosen were based on optimization tests with weighted data, but are consistent across weighted and unweighted networks.

A standard fully-connected network consists of an input layer of nodes, followed by several hidden layers connecting the inputs to the output layer. We considered networks with depth of 4, 8, or 12 hidden layers. We chose to tie the number of nodes per hidden layer to the input size with a strictly decreasing number of nodes in each successive layer. Successive layers have $\times f$ the number of the nodes of the previous layer (rounded to the nearest whole number), where we considered a fiducial value of $f=0.8$. The strategy of decreasing the number of nodes in successive layers was designed to allow the network to discard unhelpful information in the input data and only carry important or reduced information forward.

Even given the same number of hidden layers and value for $f$, the structure (and the number of free parameters) varies across the three different inputs we considered based on the size of the input information. For the $k$NN networks, the input vector has a length of 101, including the distances to the 50 neighbors, their stellar masses, and the stellar mass of the target galaxy itself. On the other hand, the cylinder counts networks take an input of length 33 to include counts in all bins and the stellar mass of the target galaxy. The combination network, which uses the information from both environmental measures, takes an input vector of 134 values.

In addition, as the stellar mass of the target object is known to be highly related to the target halo mass, we consider introducing an additional skip connection between the stellar mass input node and the layer directly before the output layer for the deeper networks (8 and 12 layers) to ensure the information contained in the stellar mass input is not discarded before the final layer.

Varying the number of hidden layers and the number of units per layer within the values considered led to little variance in the model's performance on the validation dataset ($<5$\% change in MSE) for all three inputs. The exception to this was a small subset of models, consisting mainly of deeper networks with no additional skip connection between stellar mass and output, for which the training process diverged or otherwise failed to improve upon the base SHMR. Overall, the addition of this skip connection between stellar mass and output tended to improve the performance of the deeper networks. Even with these additional connections, the deeper networks still do not show a substantial reduction in error over the shallower, 4-layer networks. Hence, we proceeded with a network structure consisting of four hidden, fully-connected layers throughout the remainder of this paper, regardless of input. Varying the base narrowing factor of $f=0.8$ by $\pm 0.1$ resulted in no significant changes in network performance. Hence, we retain the fiducial value for our final networks.

\begin{table}[h]
 \caption{Neural network parameters}
 \centering
 \begin{tabular}{lc}
  \hline
  Parameter & Values Considered\\
  \hline
  Optimization Function & \textbf{Adam}, SGD\\
  Loss Function & \textbf{MAE}, MSE\\
  Activation Function & \textbf{ReLU}, sigmoid, tanh\\
  Maximum Epochs & \textbf{50}\\
  Learning Rate &  0.01, \textbf{0.001}, 0.0001 \\
  Batch size & 32, \textbf{128}\\
  Hidden layers & \textbf{4}, 8, 12\\
  Narrowing factor ($f$) & 0.7, \textbf{0.8}, 0.9\\
  \hline \\
   \multicolumn{2}{c}{\parbox{0.9\columnwidth}{\footnotesize \textit{Notes:} Values used in final networks are shown in \textbf{bold}. Abbreviations are as follows: stochastic gradient descent (SGD), mean absolute error (MAE), mean squared error (MSE), and rectified linear unit (ReLU). Narrowing factor is defined as the fractional number of nodes in a given layer of the network when compared to the previous layer.}}
 \end{tabular}
 \label{tab:Fixed}
\end{table}

The networks were trained to minimize the loss of predicted halo masses. Training and validation losses of regression networks are typically measured via mean squared error (MSE) or mean absolute error (MAE). In this case, we chose a Mean Absolute Error (MAE) loss function. 
MAE was chosen over MSE as MAE is generally more robust to outliers and is a better choice when the data is not normally distributed or has outliers. We found that networks trained to minimize MAE converged much more quickly than networks trained with MSE. Trainable model weights were initialized from a random uniform distribution. Of the three activation functions considered (see Table \ref{tab:Fixed}), the rectified linear unit activation function (ReLU; \citealt{ReLU}) provided the best performance with the fewest training epochs. 

The two primary optimization algorithms we considered provided similar accuracy. However, the Adam optimization algorithm (\citealt{Adam2015,kingma2017adam}) provided faster training than the classic stochastic gradient descent algorithm (SGD; \citealt{SGD}). Network performance was mostly insensitive to changes in initial learning rate and batch size within the range of parameters considered. For the final models, the initial learning rate was set to 0.001, with a training batch size of 128. This information, as well as the remaining parameters of the networks described in this paper, is summarized in Table \ref{tab:Fixed}.

\begin{figure}[h]
     \centering
     \includegraphics[width=0.45\textwidth]{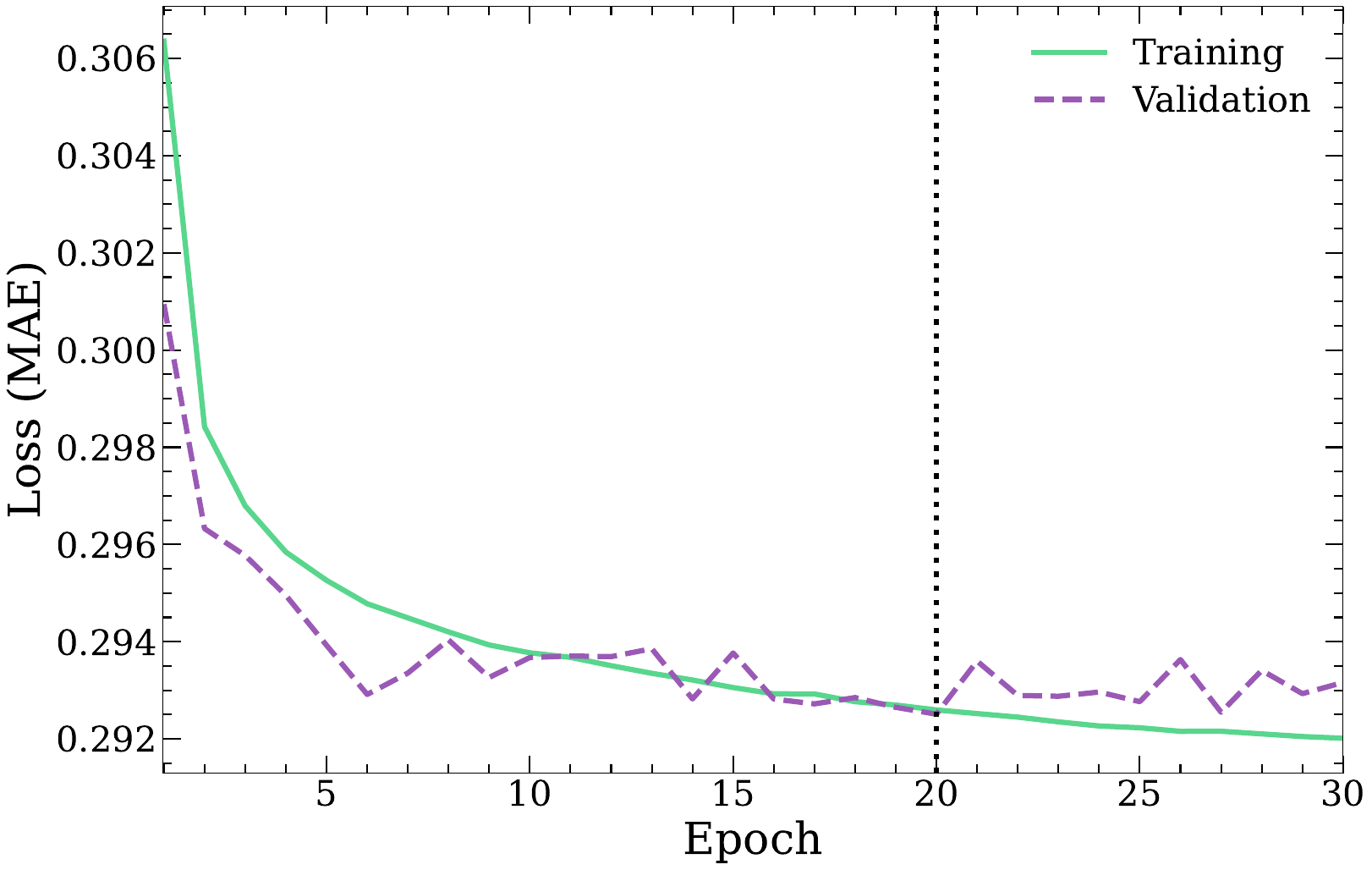}
     \caption{Evolution of the model loss during training for the unweighted $k$NN network using the finalized hyperparameter values. The loss as evaluated for the training dataset (green solid line) and the validation dataset (purple dashed line) both decrease steeply at early epochs before becoming flatter. Final model weights were taken from epoch 20 (indicated by the vertical black dotted line) as they provided the lowest validation loss.}
     \label{fig:Loss}
 \end{figure}

One potential pitfall of neural networks is lack of generalizability. Failure to train a network using a dataset representative of the intended data for the network's application may result in poor performance. This pitfall is particularly apparent if the network overfits the training set. Therefore, we allowed networks to train for up to 50 epochs but imposed an early stopping criterion to reduce the potential overfitting of the training dataset. After each epoch, the validation dataset's loss was assessed, and if there was no improvement within 10 epochs (i.e., a patience of 10), the training was halted. A warm-up period of ten epochs was implemented to prevent training from being stopped before a solution is found. The final model weights were then selected from the epoch with the best validation loss score. Figure \ref{fig:Loss} shows the evolution of training and validation loss with epoch during the training period of the unweighted $k$NN network. The training for this network was stopped at 30 epochs, as the validation loss had not decreased below the 20 epochs value. In the following section, we analyze the results of networks trained with and without weights as applied to the test data.

\section{Results}\label{sec:Results}

In this section, we compare the performance of the nearest neighbors, cylinder counts, and combined models when applied to the \textit{Bolshoi-Planck} test data. We consider the performance of the models over (1) the whole dataset, (2) bins in halo mass, and (3) central and satellite galaxies separately. Errors are compared against the scatter in the SHMR as described in Section \ref{subsec:SM_Alone}. Section \ref{subsec:kNN} describes the performance of the fifty nearest neighbors networks, while Section \ref{subsec:kNN_Features} explores the impact of removing the data from more distant neighbors. Similarly, Sections \ref{subsec:CoC_Results} and \ref{subsec:CoC_FI} provide an analysis of the cylinder counts networks and the relative importance of the different features used therein. Lastly, Section \ref{subsec:Combo} covers the networks which take a combination of the two environmental measures. For each model, the loss values shown represent the root mean squared error (RMSE) in the predicted halo mass compared to the true values from the simulation unless otherwise indicated. All reported uncertainties, including error bars and shaded regions, correspond to 68\% confidence intervals.

\begin{figure}[h]
     \centering
     \includegraphics[width=0.45\textwidth]{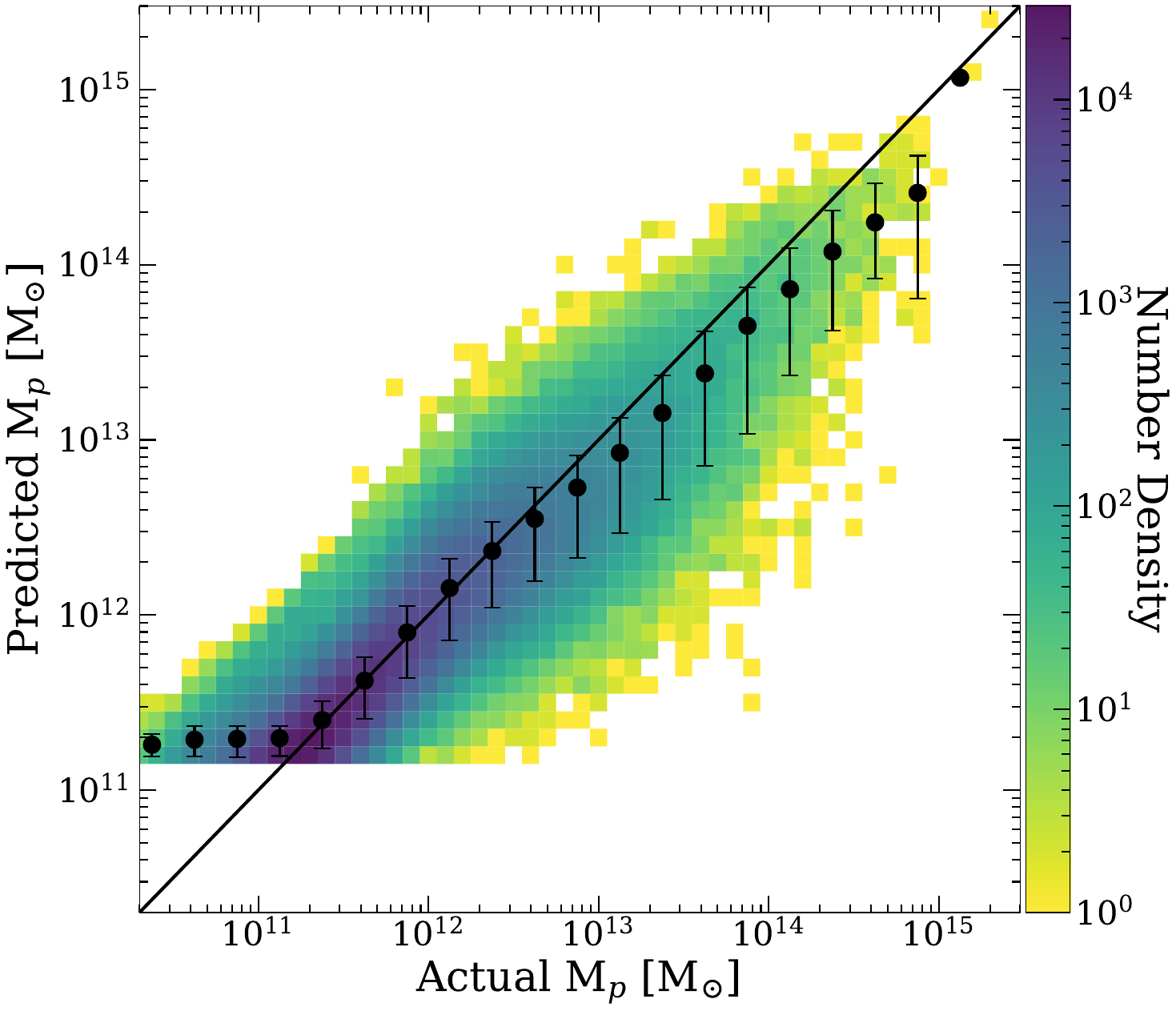}
     \caption{Predicted host-halo masses for \textit{Bolshoi-Planck} galaxies from interpolation of the average stellar-mass halo-mass relation (SHMR) in SMDPL are plotted against true halo mass values from the simulation. Darker regions in the 2D-histogram of the data distribution correspond to the more dense regions of prediction versus actual values space. Black points represent the median predictions of halo mass by bin in true halo mass, with error bars showing the 68\% scatter of the data in the bin.}
     \label{fig:SM1}
\end{figure}

\subsection{Stellar Mass Alone}\label{subsec:SM_Alone}

There exists a well-constrained relationship between observed stellar mass (or, in practice, cumulative number density) and the peak mass of the host dark matter halo \citep{Wechsler_2018}. To provide a baseline prediction of halo mass from stellar mass alone, we first found the average SHMR from SMDPL (effectively, the average halo mass as a function of observed stellar mass from the \textsc{UniverseMachine}). We then used this average relationship to assign masses to halos in \textit{Bolshoi-Planck}. The results of this method are shown in Figure \ref{fig:SM1}.

While scatter in the SHMR is often reported as the scatter in stellar mass at fixed halo mass, as we are estimating halo masses from known stellar masses, we are interested in the scatter in halo mass at fixed stellar mass. For stellar masses of $10^9 - 10^{10} \Msun$ (corresponding on average to halo masses $10^{10.5}-10^{11.5} \Msun$) the one-sigma scatter in halo masses within SMDPL is $\lesssim 0.2$ dex. As stellar mass increases, so does the scatter in stellar mass,  reaching more than $0.5$ dex at $\sim 10^{11.5} \Msun$. Figure \ref{fig:SM1} shows how this results in increasing scatter in halo mass estimationfor $M_p > 10^{12} \Msun$.

\begin{figure}
     \centering
     \includegraphics[width=0.45\textwidth]{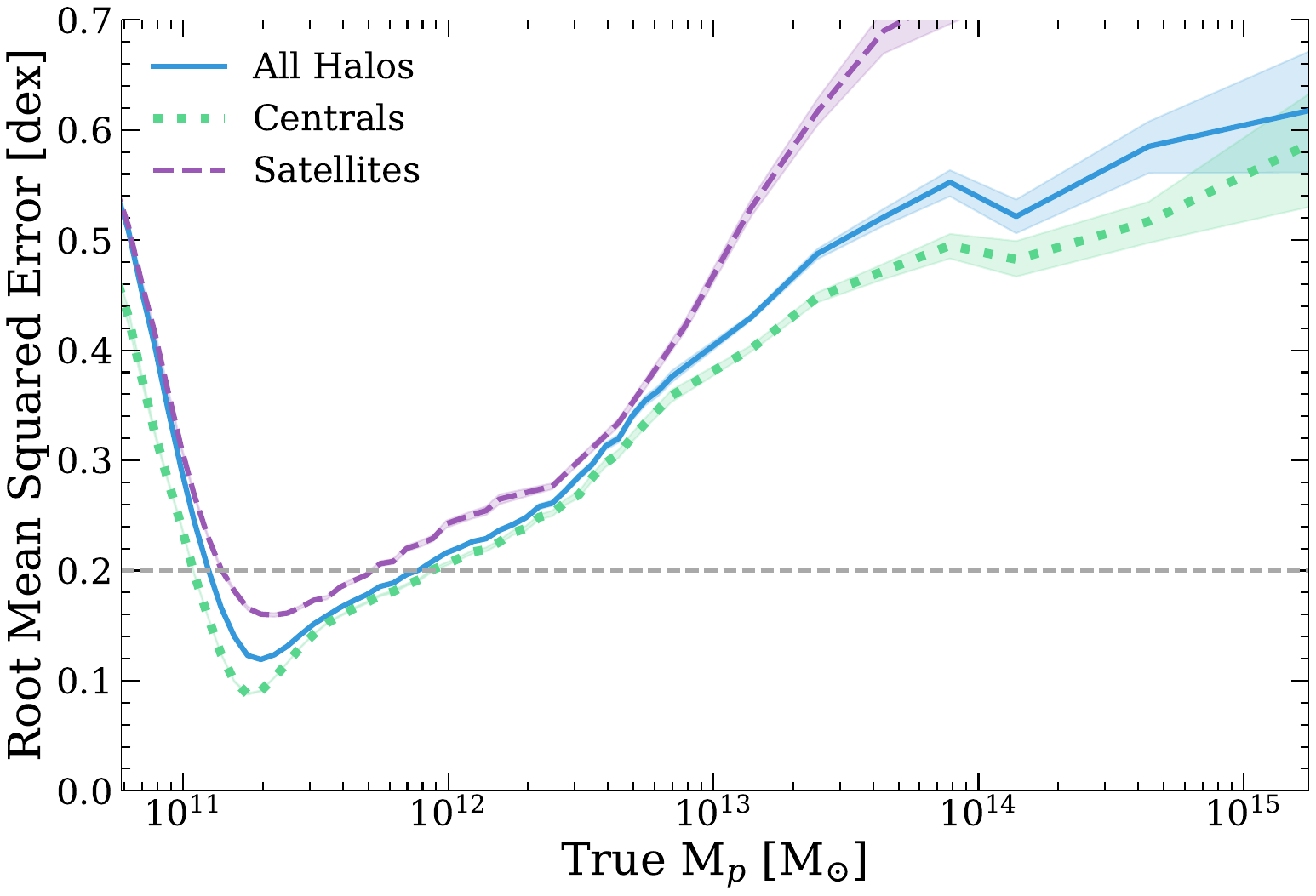}
     \caption{Error (RMSE in dex) in halo mass estimated via direct interpolation of the SHMR is plotted as a function of true halo mass. This figure and subsequent figures of this format show the accuracy of the given method (in this case interpolation from stellar mass) for a given true peak halo mass. The population is split into centrals (green dotted line) and satellites (purple dashed line). The performance over the whole population is shown by the blue solid line. Shaded regions show 68\% scatter as estimated from bootstrap resampling. Excluding the lowest-mass region where the error is driven by sample selection bias (see Section \ref{subsec:SM_Alone} for further discussion), error tends to increase with halo mass. A horizontal dashed line shows 0.2 dex accuracy, which is the relevant accuracy limit for applications to galaxy formation. This demonstrates that the interpolation method has a substantially lower accuracy at high halo masses than at low halo masses, with central halo predictions being more accurate than satellites.}
     \label{fig:SM_CvS}
\end{figure}

Figure \ref{fig:SM_CvS} shows the average error in predicted halo mass as a function of true halo mass (blue solid line). The upturn in the loss for $M_p < 10^{11} \Msun$ is primarily the result of a sample selection effect. As we excluded galaxies with stellar masses less than $10^9 \Msun$ from our sample, we preferentially selected low-mass halos hosting over-massive galaxies for their size. Excluding this, the overall trend is towards higher prediction errors for higher mass halos. For the interpolation method, only a small range of true halo masses ($M_p = 10^{11}-10^{12} \Msun$) is predicted to within the desired accuracy of 0.2 dex.

\begin{figure}[h]
     \centering
     \includegraphics[width=0.4\textwidth]{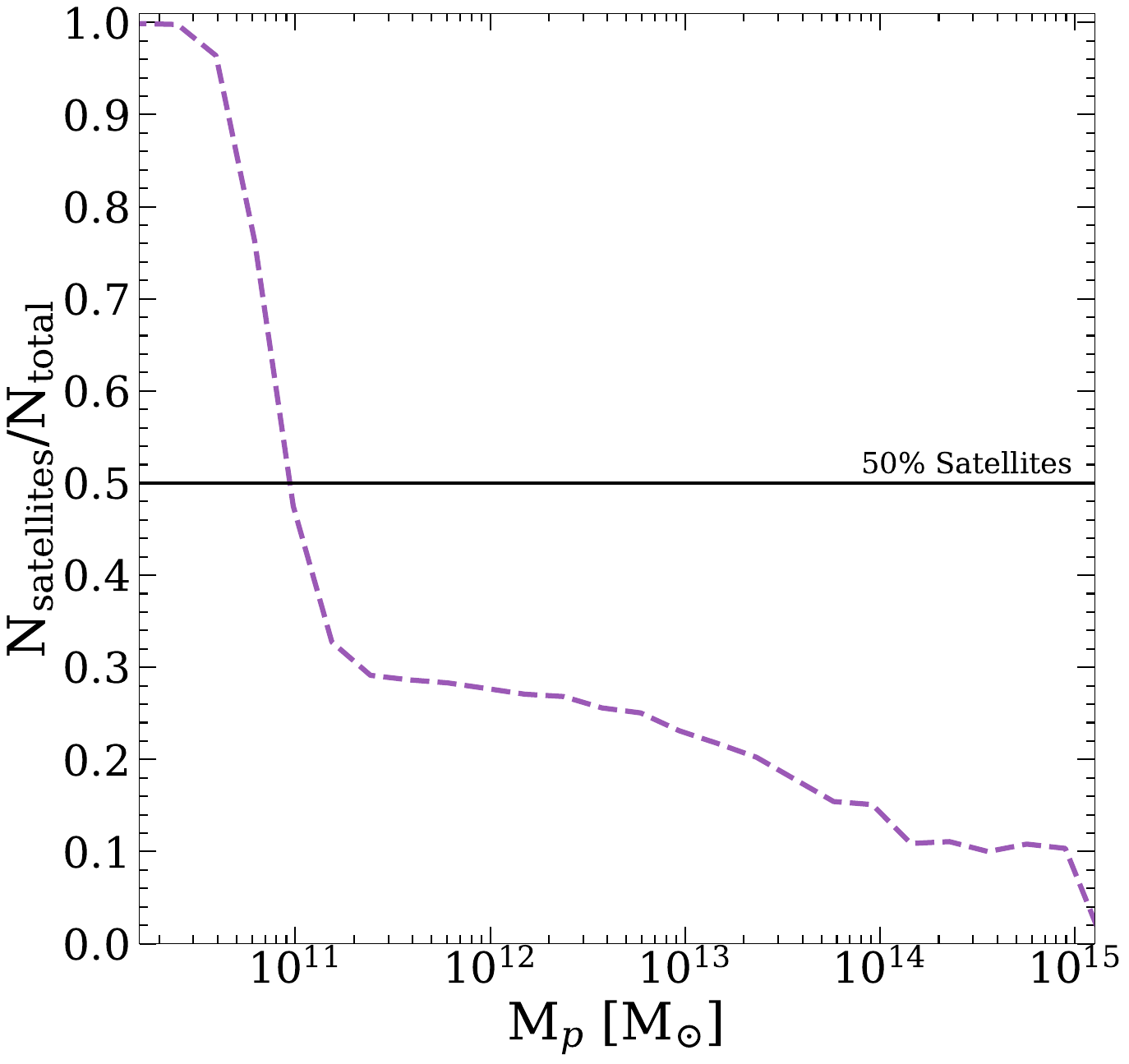}
     \caption{The fraction of galaxies that are satellites in bins of peak halo mass. Satellites make up the majority of galaxies with host halos below $M_p \sim 10^{11} \Msun$, due to the stellar mass cut at $10^{9} \Msun$, which eliminates objects with a typical SHMR for centrals in this mass regime (see text for further explanation).}
     \label{fig:Qsats}
 \end{figure}

Another consideration regarding interpolation from stellar mass is the different behaviors of central versus satellite halos. A large portion of our sample, particularly at the low halo mass end, is composed of satellites. Figure \ref{fig:Qsats} shows the fraction of satellites (purple dashed line) in the SMDPL box as a function of peak halo mass. The sharp upturn in the satellite fraction at low mass is a reflection of the higher stellar mass – halo mass ratios in subhalos in combination with a fixed stellar mass cut. This trend can be seen in Figure \ref{fig:MassDist} where the distribution of satellite halos (green filled contours) is shifted towards higher stellar masses for a given halo mass compared to the central halos (black contour lines). This trend occurs because galaxies in subhalos continue forming stars even after their halos stop accreting matter, leading to higher stellar mass to peak halo mass ratios than centrals \citep[e.g.,][]{UM}.  With a fixed stellar mass cut, the galaxies selected at the lowest halo masses will have the highest ratios of stellar mass to halo mass, which means that primarily galaxies in subhalos will be selected. 

To demonstrate this, the full halo population (solid blue) in Figure \ref{fig:SM_CvS} is divided into central (green dotted) and satellite (purple dashed) halos. A comparison of these lines shows that the stellar mass is sufficient for predicting the mass of both central and satellite halos below $M_p = 10^{12} \Msun$. At higher halo masses, stellar mass alone provides more information about central halos than satellite halos, leading to higher errors for satellites. Overall, the inclusion of satellite halos drives up the error in predicted halo mass from $\lesssim 0.6$ dex for central halos alone across the full mass range.

We preform the same test separating centrals and satellites on the three network types discussed below. The results of all three are highly similar and so discussion is reserved to Section \ref{subsec:Combo}.

 \subsection{Nearest Neighbors Results}\label{subsec:kNN}

\begin{figure}
     \centering
     \includegraphics[width=0.45\textwidth]{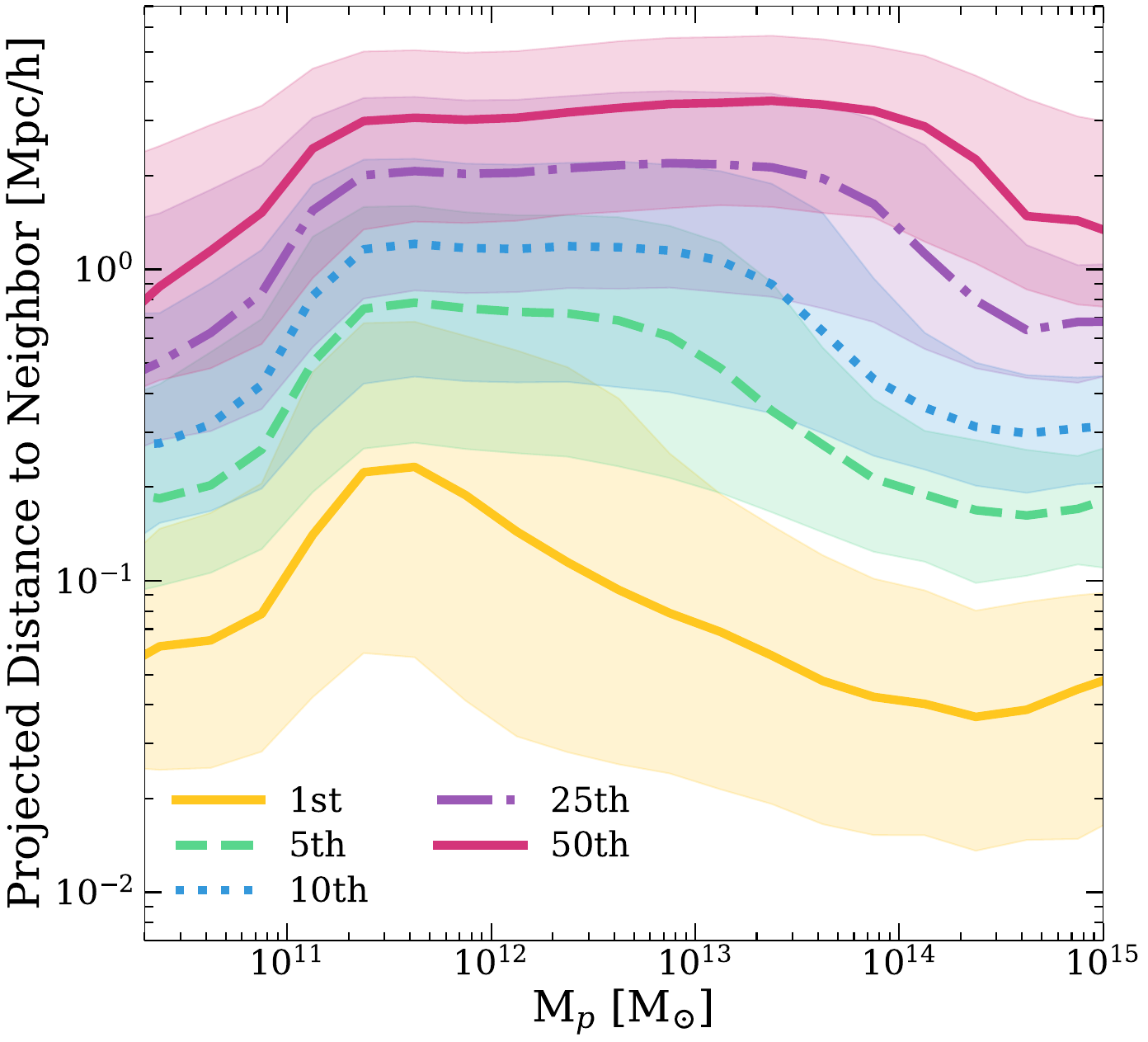}
     \caption{Median distance to $k$th nearest neighbor vs. peak halo mass. Each line represents the median distance as evaluated in 0.1 dex bins of halo mass for different values of $k$. The shaded regions indicate the 16-84th percentile regions.}
     \label{fig:Neighbors}
 \end{figure}

The first source of environmental information we consider is the $k$NN distances. As shown in Figure \ref{fig:Neighbors}, the average projected distance to the $k$th nearest neighbor depends on halo mass. Halos with peak masses of $10^{11} - 10^{12} \Msun$ tend to be found in low-density environments (large distance to the $k$th nearest neighbor) while halos at the group and cluster mass scales ($M_p > 10^{12.5} \Msun$) are found in denser galaxy environments (small distance to the $k$th nearest neighbor). Below $M_p \sim 10^{11} \Msun$, the average distance to neighbors decreases again. In this regime, nearly all halos ($\sim 99\%$) are satellites of massive halos (Fig \ref{fig:Qsats}), and thus they are also found in high-density environments  (i.e., they inherit a high-density environment from the nearby central halo they orbit).

From Figure \ref{fig:Neighbors}, we can also see that the dependence of $k$NN distance on halo mass varies with the value of $k$, suggesting that certain values of $k$ may be more sensitive to halo masses in different regimes. For example, the distribution of distances to the 50th nearest neighbor does not considerably change between halo masses of $10^{11} \Msun$ and $10^{14} \Msun$, so we can expect that this value will not be a useful probe in this regime. However, it may be helpful for distinguishing halo masses between $10^{14} \Msun$ and $10^{15} \Msun$, where the slope of the relationship is steeper. 

\begin{figure}
     \centering
     \includegraphics[width=0.45\textwidth]{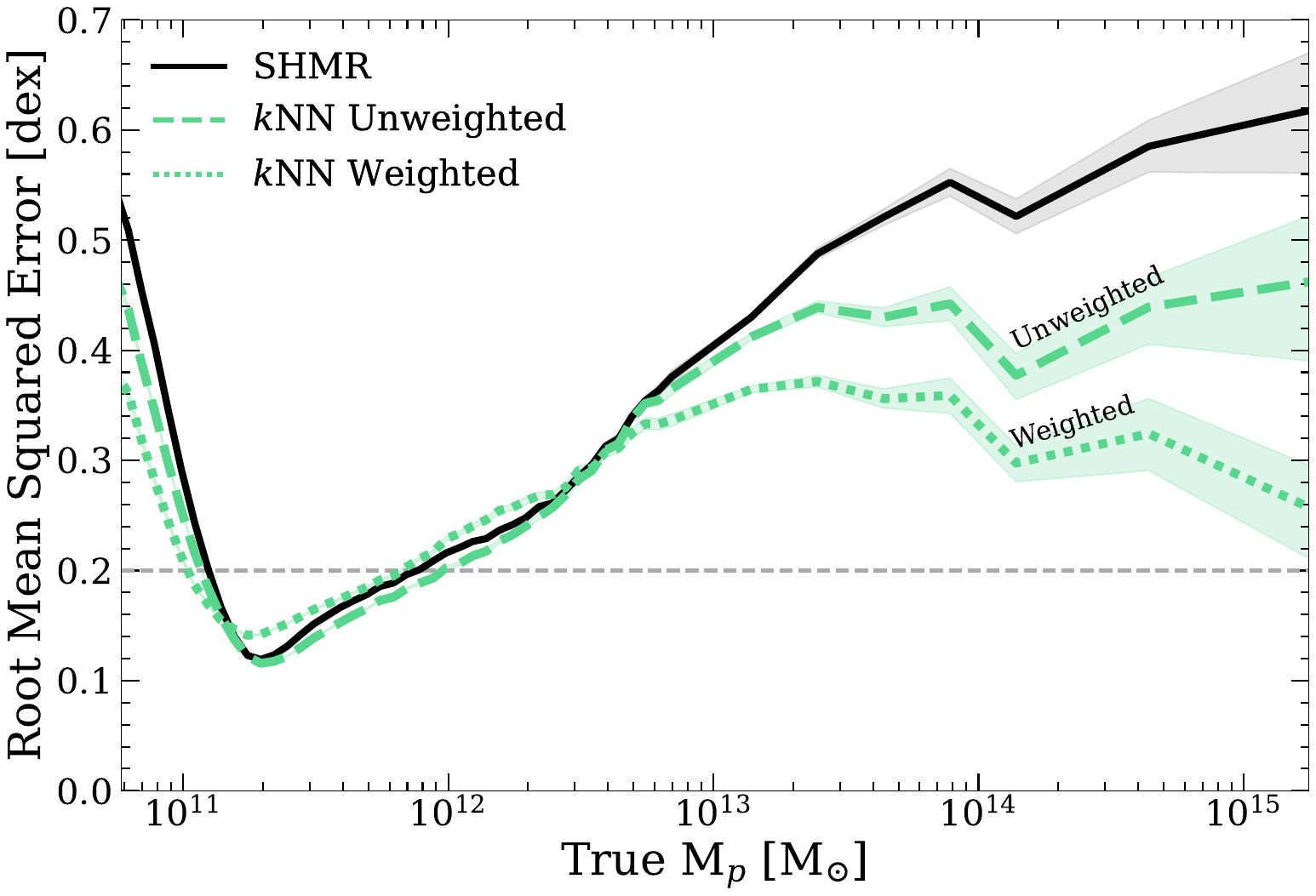}
     \caption{Loss (RMSE in dex) is plotted as a function of halo mass for the unweighted (green dashed line) and weighted (green dotted line) $k$NN networks. The performance of the SMHR interpolation from Section \ref{subsec:SM_Alone} is shown by the black solid line for comparison. Shaded regions show 1$\sigma$ errors as estimated from bootstrap resampling. A horizontal dashed line shows 0.2 dex accuracy for reference.}
     \label{fig:kNN1}
\end{figure}

When training with no prior weighting of the data, the best overall RMSE achieved was 0.19 dex. As shown by Figure \ref{fig:kNN1}, the losses from the network (green dashed line) have a similar shape as the interpolation from the SHMR (black solid line) for halo masses below $10^{12.3} \Msun$. Above this threshold, the network outperforms SHMR interpolation with errors tending to decrease at higher halo masses and remaining between $\sim$ 0.3 and 0.4 dex, compared to the median error of 0.6 dex found by the SHMR interpolation for $10^{14} - 10^{15} \Msun$ halos. 

We also consider a network trained on weighted data (as described in Section \ref{subsec:Preprocess}), which outperforms both the stellar mass interpolation and the unweighted network at halo masses > $10^{12.1} \Msun$, with a sacrifice in accuracy at lower masses (green dotted line in Figure \ref{fig:kNN1}) and a slightly lower overall accuracy of 0.20 dex. Hence, in applications where the higher-mass end is more important, the weighted network would be more relevant, and vice versa for lower masses.

In both the unweighted and weighted network results, the median error in the prediction of the networks for halo masses below $\sim 10^{12} \Msun$ does not substantially improve on the stellar mass-only prediction. In the higher mass regime, the neighbor information is more helpful since halos in this regime tend to have multiple satellites, and thus the neighbor information is likely more sensitive to the target halo's mass than to the large-scale environment. In particular, in the galaxy group and cluster regime ($M_p \gtrsim 10^{12.5} \Msun$), the number of satellites, as indicated by the neighbor density, is expected to be strongly correlated with halo mass \citep{Muldrew2012}.

\subsection{Nearest Neighbors Feature Importance}\label{subsec:kNN_Features}

\begin{figure}
     \centering
     \includegraphics[width=0.45\textwidth]{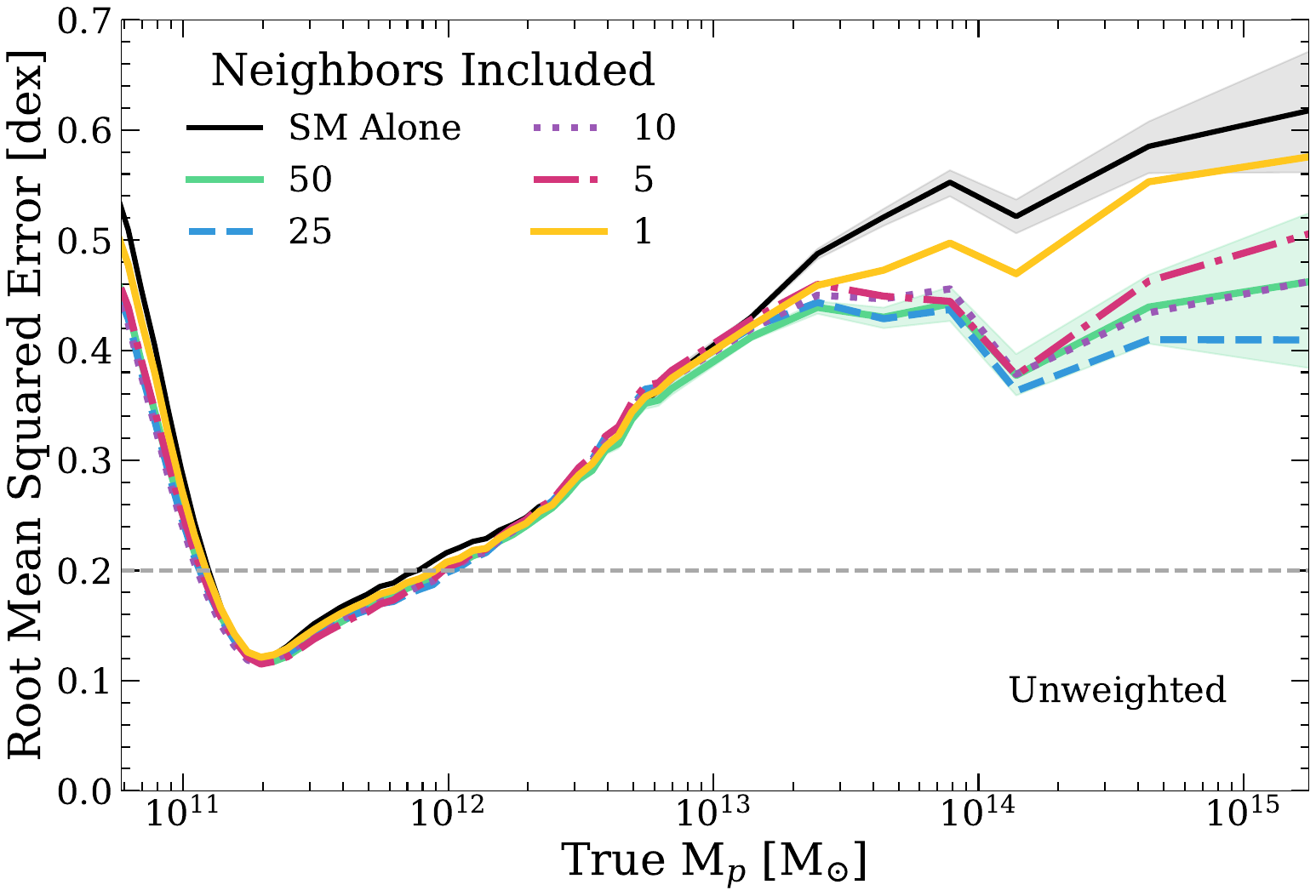}
     \includegraphics[width=0.45\textwidth]{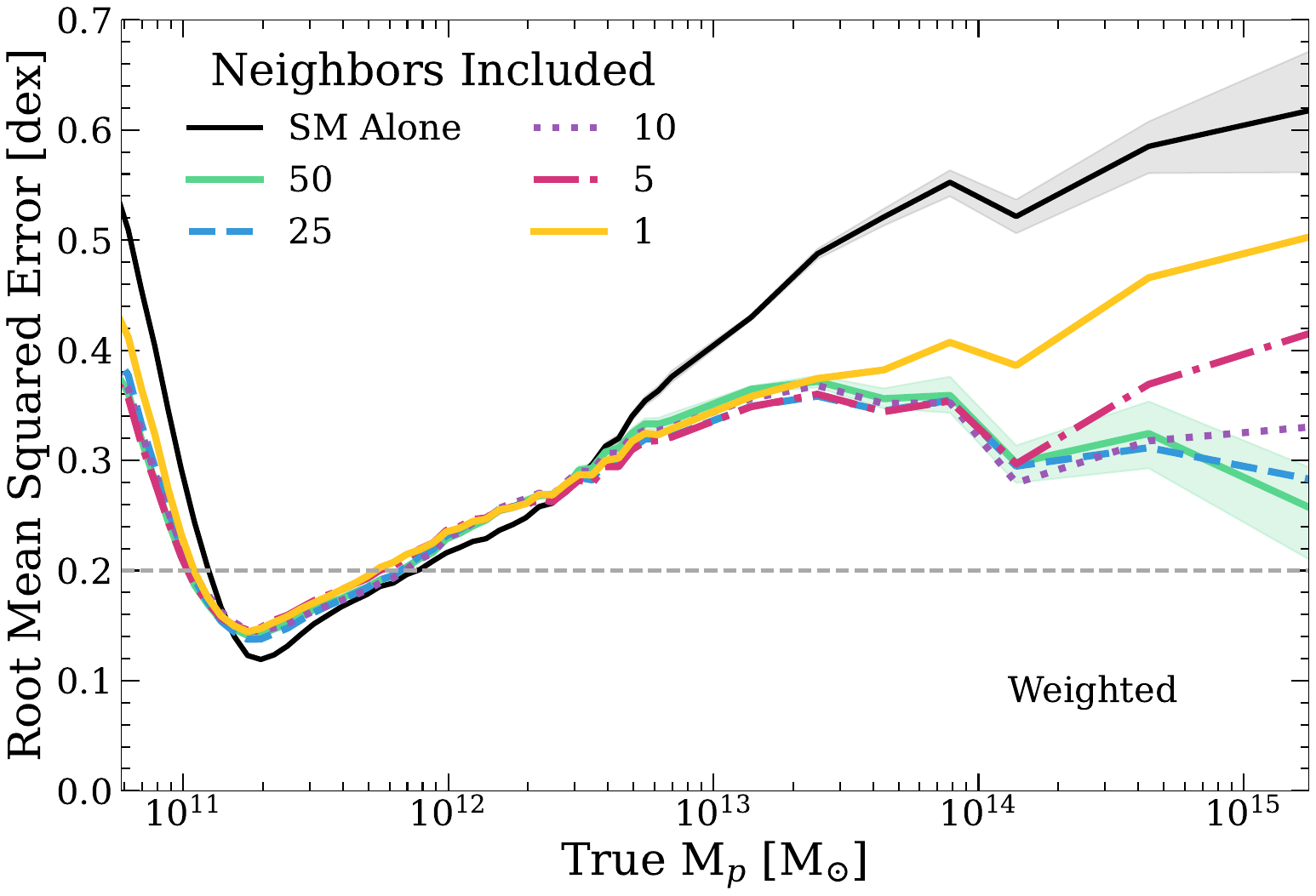}
     \caption{Loss (RMSE in dex) of unweighted (top) and weighted (bottom) networks is plotted as a function of halo mass for the four unweighted nearest neighbors networks trained on 1, 5, 10, 25, and 50 neighbors (pink, purple, blue, green, and yellow respectively). The black solid line shows the loss from the SHMR interpolation (i.e., using information from stellar mass alone) for comparison. The shaded green regions show the 68\% scatter in the loss for the full network as estimated from bootstrap resampling. The error regions of the other lines are excluded for clarity but are of similar magnitudes to the ones shown. A horizontal dashed line shows 0.2 dex accuracy for reference.}
     \label{fig:kNN2}
 \end{figure}

\begin{figure*}
     \centering
     \includegraphics[width=\textwidth]{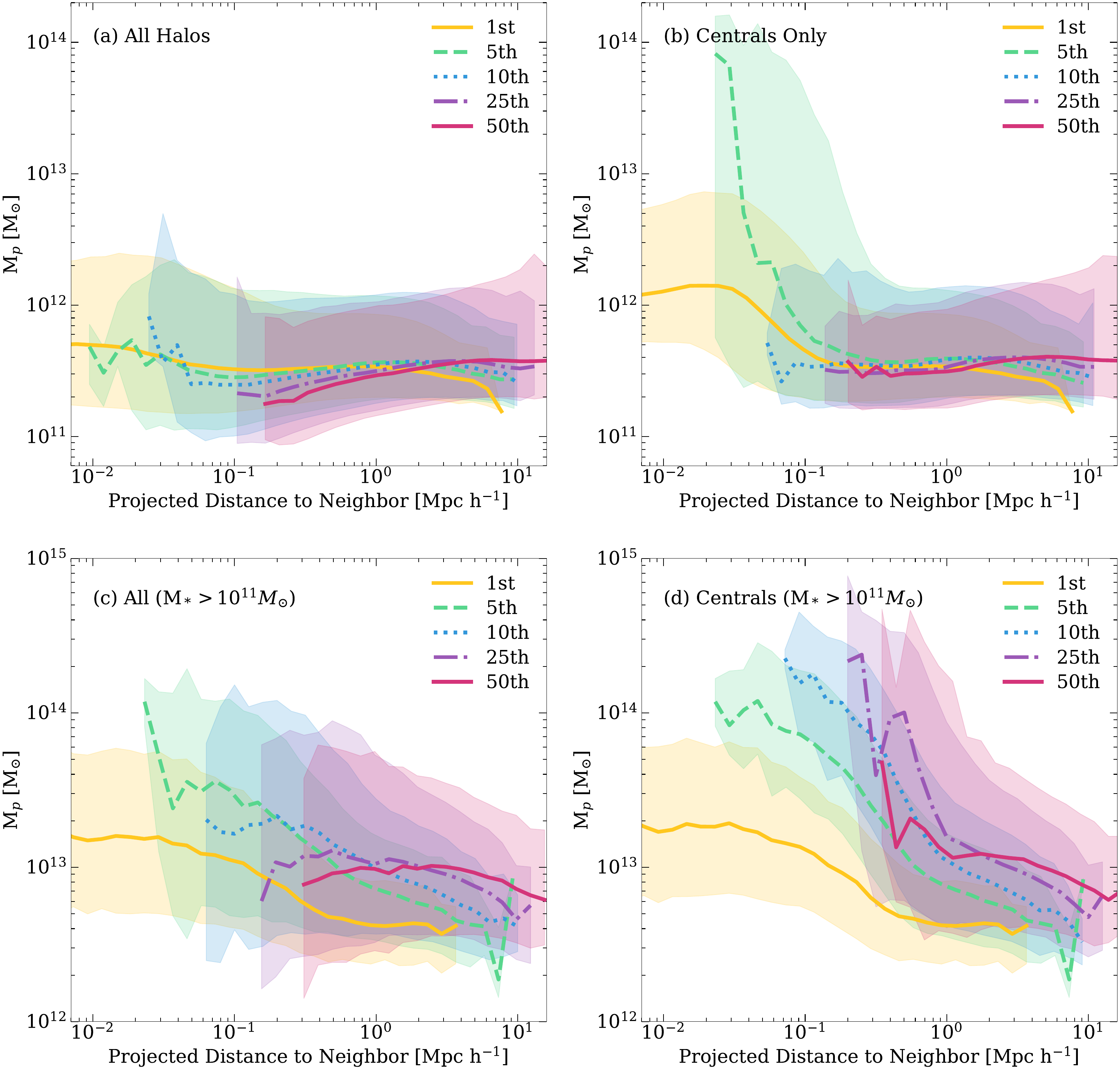}
     \caption{Average peak halo mass in bins of distance to the $k$th nearest neighbor. Each line represents the median halo mass for a different value of k, with the shaded regions indicating the 16-84 percentile region. The top row includes the full mass range for all halos (a) and for centrals only (b). The bottom row limits the population to halos hosting galaxies with M$_* > 10^{11} \Msun$ for all halos (c) and for centrals only (d).}
     \label{fig:Dists2}
\end{figure*}

To decipher how the given information is used by the networks, we attempted to isolate the impact of certain input features on the network predictions. In the case of an average mass halo, we may expect that the first several neighbors provide information about the halo’s satellites, while the remaining neighbors probe a larger region out to several Mpc, which may or may not provide additional information about the halo’s mass. On the other hand, given the choice of a fixed number of neighbors, including neighbors out to fifty (or beyond) could potentially be necessary to probe the full satellite populations of the most massive halos.

To determine the relevance of including more distant neighbors, we performed a process where information about neighbors beyond a given value of $k$ was masked (i.e., the feature was replaced with the value zero) and the network re-trained. Each iteration was executed with the same network structure and hyperparameters as for the full 50 neighbors case. Figure \ref{fig:kNN2} shows the resulting errors in predictions, with each colored line representing a different number of neighbors included. 

Masking neighbors did not result in changes in network performance for $M_p \lesssim 10^{12.5}$. This is to be expected, as the performances of the full networks are not distinguishable from the SHMR interpolation in this regime. The impact of masking neighbors only becomes apparent at higher masses. In the unweighted case, the one-neighbor network does not perform as well as the networks provided with more neighbors for $M_p \gtrsim 10^{13} \Msun$. However, there is no significant difference between the performances of the $5\leq k \leq 50$ neighbor networks, all falling within the confidence interval of the 50-neighbor network. 

For the weighted networks, we found that up to halo masses of $\sim 10^{13} \Msun$ and $\sim 10^{14} \Msun$, predictions based on one neighbor (yellow line) and five neighbors (pink line), respectively, are as accurate as predictions based on larger neighbor numbers, diverging at higher masses. There is no substantial difference in accuracy between the ten, twenty-five, and fifty-neighbor networks. Except for the one-neighbor case, each weighted network achieves an average loss of $\leq 0.20$ dex. Including more neighbors beyond five does not significantly alter the performance of the overall network as constructed in the weighted cases. When evaluating loss as a function of halo mass in Figure \ref{fig:kNN2}, including five neighbors is a clear improvement on the one-neighbor case at the high mass end.  Including ten neighbors may result in slight improvements for clusters with $M_p \gtrsim 10^{14}\Msun$. There is no clear improvement gained in moving to more than ten neighbors.

The process is not extended beyond fifty neighbors, due to the lack of clear improvement in including more than ten neighbors. To better understand why the more distant neighbors are not more informative about halo mass, we considered the average halo mass as a function of distance to the $k$th nearest neighbor (Figure \ref{fig:Dists2}). If we consider all halos (a), no one value for $k$ stands out as particularly informative, however, when limiting to centrals only (b), the 5th neighbor's distance stands out as having the largest variance with halo mass. 

If we limit our analysis to the high mass end ($M_{*} > 10^{11} \Msun$) where the inclusion of neighbor information was found to improve the network's performance, the 5th neighbor still stands out as the most relevant (Figure \ref{fig:Dists2}, panel c). However, when limited to high mass centrals (panel d), there is an apparent trend in halo mass with distance for $5 \leq k \leq 25$. This suggests, as expected, that more distant neighbors also carry information about the most massive central halos. However, it is also possible that some information at higher values of k is redundant. The differing performance of the network on central and satellite halos is discussed further in Section \ref{subsec:Combo}.

\begin{figure}
     \centering
     \includegraphics[width=0.45\textwidth]{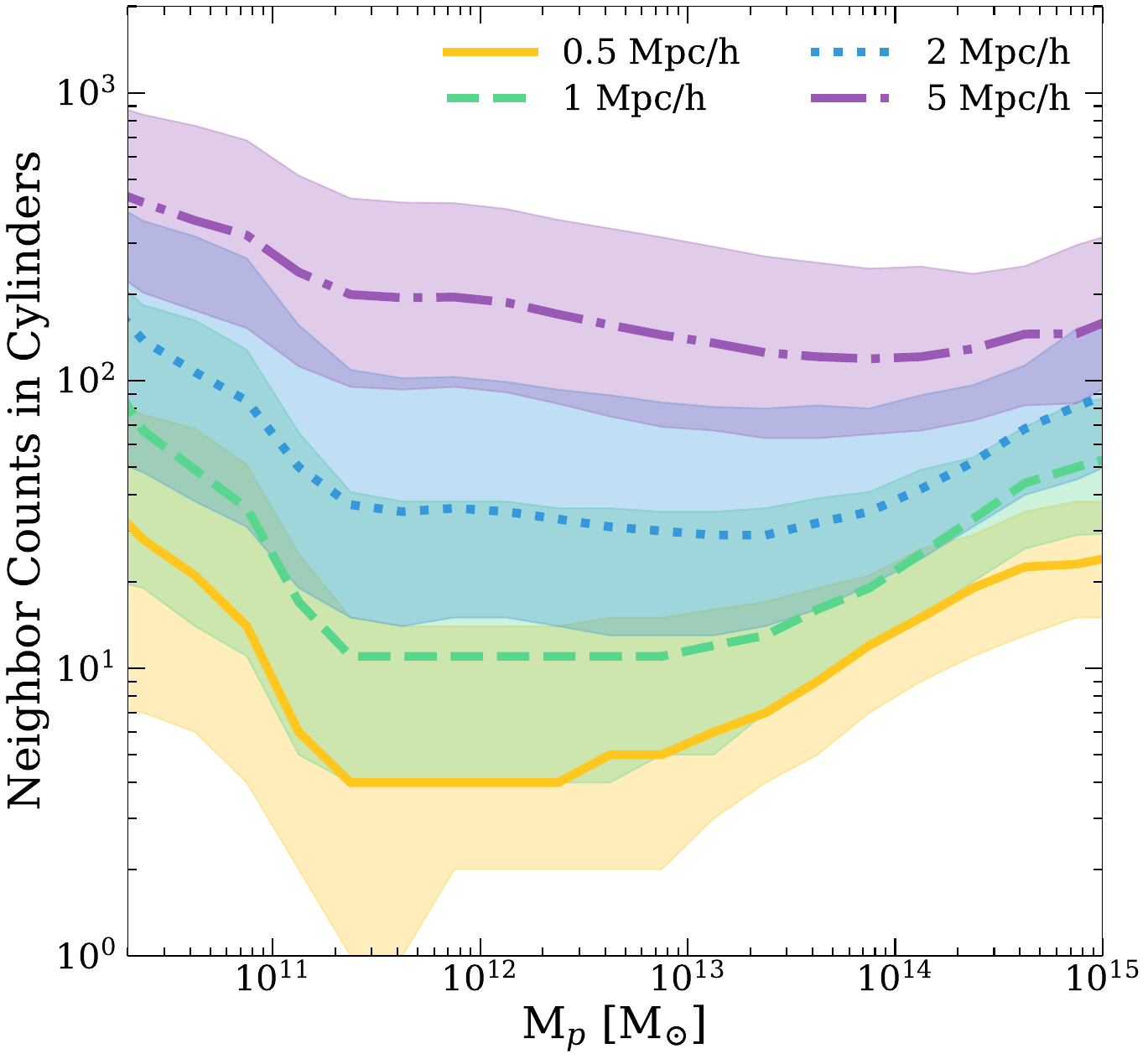}
     \caption{Median number of neighboring galaxies falling within a $\pm$250 km s$^{-1}$ depth cylinder as a function of halo mass evaluated in 0.1 dex mass bins. The shaded regions indicate the 16-84th percentile regions.}
     \label{fig:Cylinders}
 \end{figure}

\subsection{Counts in Cylinders Results}\label{subsec:CoC_Results}

\begin{figure}
     \centering
     \includegraphics[width=0.45\textwidth]{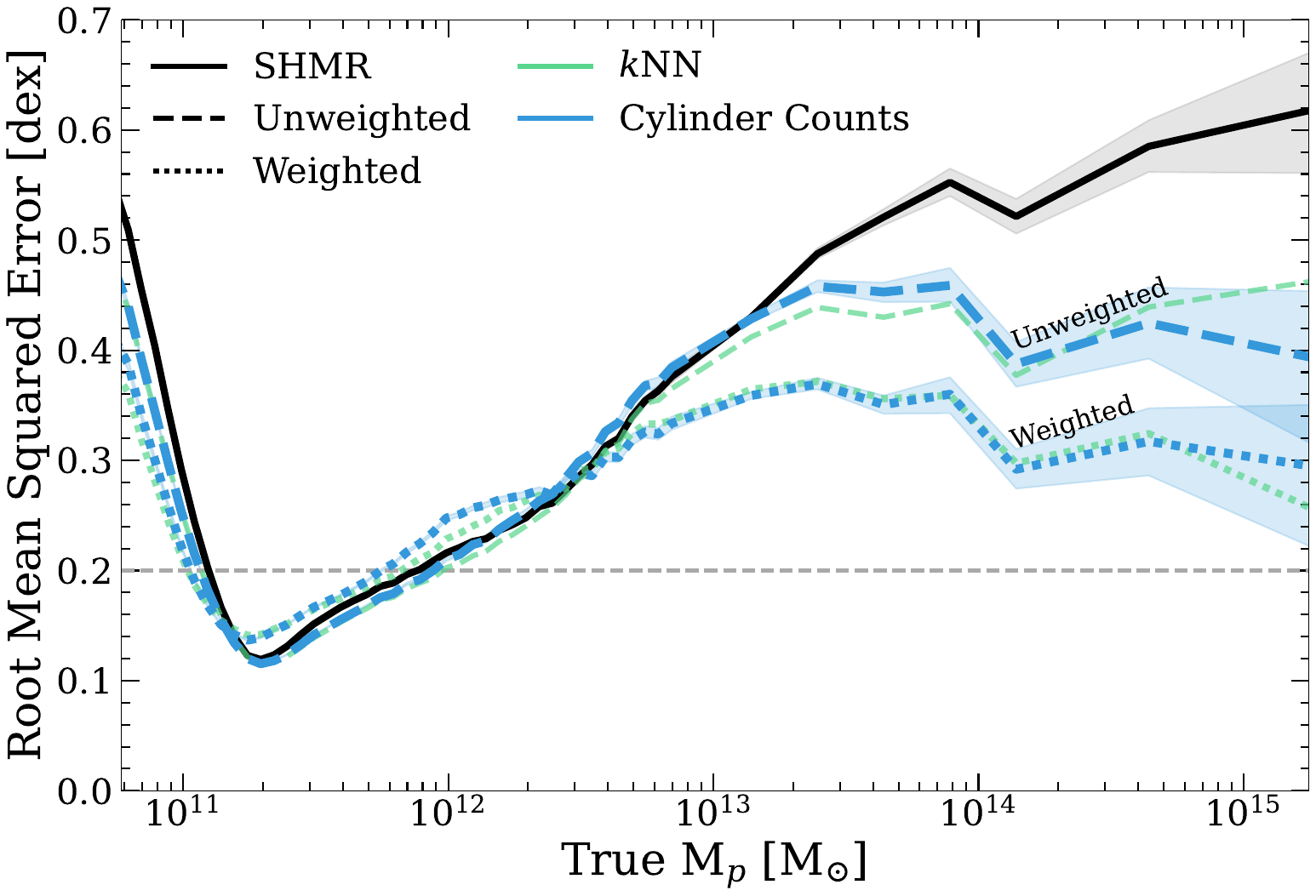}
     \caption{Loss (RMSE in dex) is plotted as a function of halo mass for the unweighted (blue dashed line) and weighted (blue dotted line) cylinder networks. The performance of the $k$NN networks (green) and the SMHR interpolation (black) are shown for comparison. Shaded regions show 1$\sigma$ errors as estimated from bootstrap resampling. These are excluded for the $k$NN networks for clarity, but are of similar magnitude to the cylinder counts error regions. Overall, the performance of the cylinder counts networks are highly similar to their nearest neighbor counterparts. A horizontal dashed line shows 0.2 dex accuracy for reference.}
     \label{fig:Cyl1}
 \end{figure}
 
We also consider the alternative environmental measure of counts in cylinders as defined in Section \ref{subsubsec:CoC}. Figure \ref{fig:Cylinders} shows how the median value for counts in cylinders evolves with halo mass for different cylinder sizes. Given the similarity in the information provided to the two networks, we expected the network trained on counts in cylinders to have a performance highly similar to the nearest neighbors network. This similarity is evident in Figure \ref{fig:Cyl1}, where the loss values for cylinder counts (blue) and $k$NN (green) are similar. We anticipated that the cylinder counts network could have an advantage for the highest mass clusters, where the 5 $h^{-1}$Mpc radius bin, as well as the inclusion of redshift separations up to 2000 km s$^{-1}$, might cover more of a cluster's satellites than the 50 nearest neighbors measure. Additionally, the finer redshift binning might allow the network to better separate nearby galaxies from those with low projected distances but with large velocity offsets. On the other hand, by using counts instead of masses (or some proxy of masses), there is other information lost. For example, with counts alone, the network would not have a way to know if one of the nearby neighbors is more massive than the target galaxy, which might otherwise allow the network to determine whether the target galaxy is a central or a satellite.

For the counts in cylinders, we used a dense network with 33 inputs and 4 hidden layers. The overall loss was 0.20 dex for both the unweighted network and weighted networks. Figure \ref{fig:Cyl1} shows the performance of the unweighted (dashed blue line) and weighted (dotted blue line) networks compared to stellar mass alone (black solid line) and the full fifty neighbor networks described in section \ref{subsec:kNN} (green dashed and dotted lines). 

There is little difference in performance between the nearest neighbor networks and the cylinder networks. In both the unweighted and weighted cases, the lines denoting the nearest neighbor network's loss fall within the shaded error region for the corresponding cylinder network. The similarity in performance between the $k$NN networks and the counts in cylinders networks suggests that the neural networks are not able to extract more information from one environmental measure than the other. 

\begin{figure*}[h]
     \centering
     \includegraphics[width=0.9\textwidth]{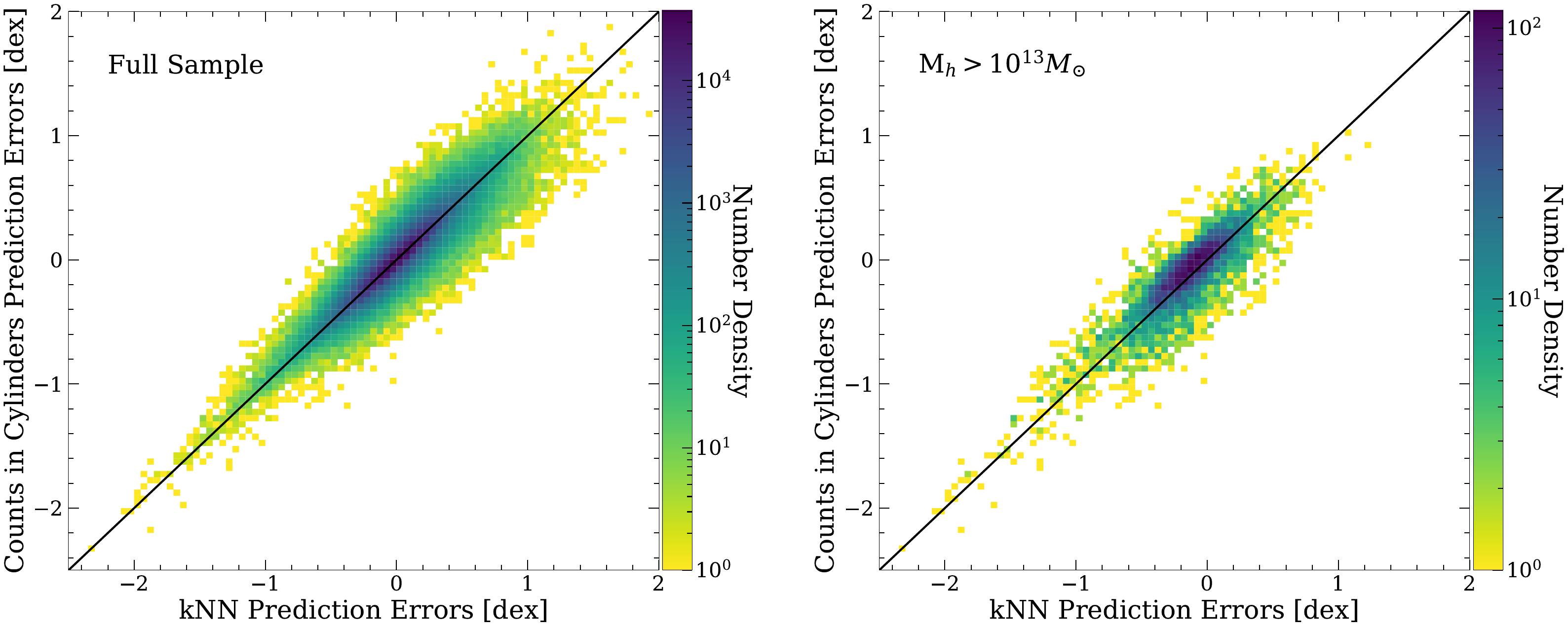}
     \caption{Difference between predicted and actual values of halo mass in dex from the full weighted counts in the cylinders network versus the full weighted $k$NN network. The number density of halos is shown by the color bar. Left: Offset between predicted and actual values for the full test sample in log-space. The two networks' errors are strongly correlated, with a Pearson correlation coefficient of 0.93. Right: Offsets when limited to the high-mass end ($M_h > 10^{13} \Msun$). At lower halo masses, predictions are driven primarily by stellar mass. This second plot shows the similarity between the two networks ($\rho = 0.87$) when applied outside the stellar mass dominated prediction range.}
     \label{fig:Combo1}
\end{figure*}

We can additionally ascertain whether the networks are extracting the same information by comparing the mass-prediction errors for individual galaxies. Figure \ref{fig:Combo1} shows the errors in estimated halo masses from the weighted counts in cylinders network versus those from the weighted $k$NN network. The predictions of the two networks are highly similar, even when limitted to the high-mass end, suggesting that they are likely extracting the same information from the different environmental measures. To further evaluate this, we considered a network provided with both the $k$NN and cylinder counts information in section \ref{subsec:Combo}.

\begin{figure*}
     \centering
     \includegraphics[width=0.9\textwidth]{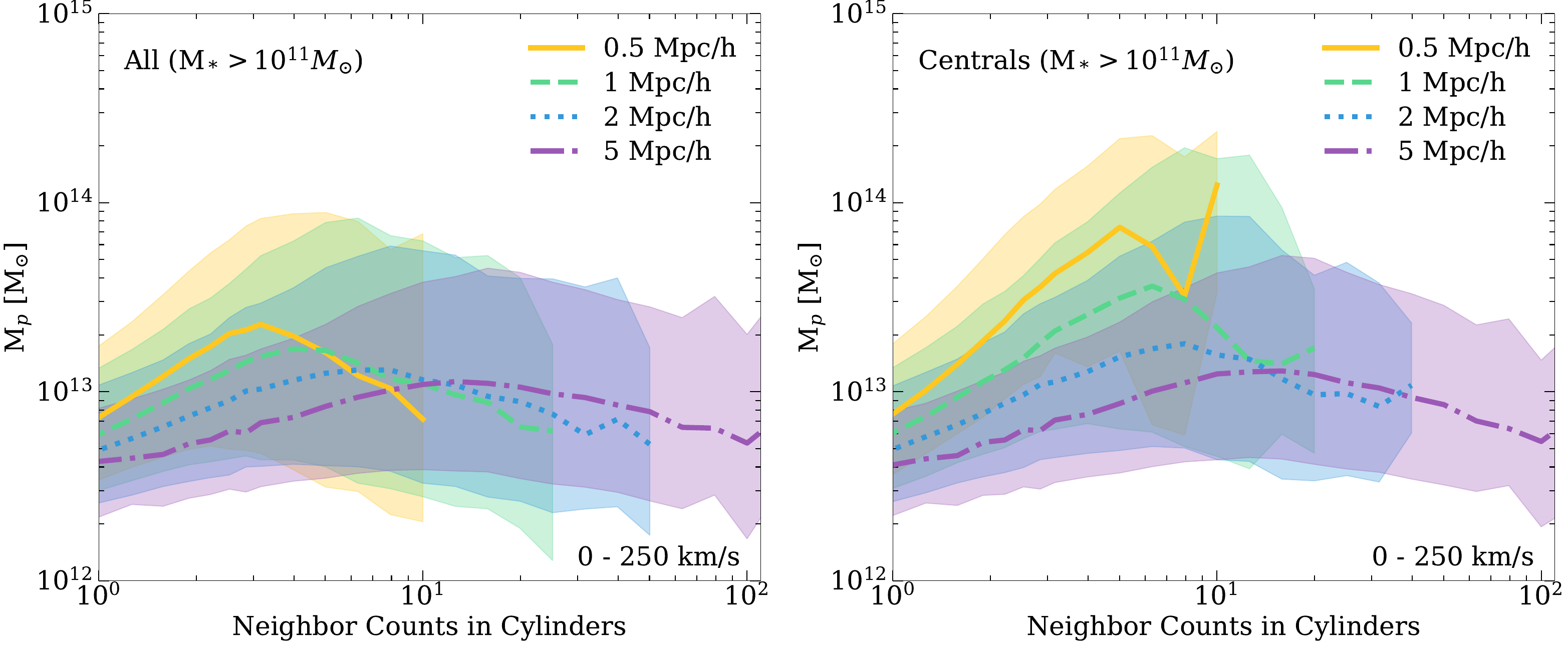}
     \caption{Average peak halo mass in bins of neighbor counts. Each line represents the median halo mass for a given number of counts within the specified radius, with the shaded regions indicating the 16-84 percentile region. Only counts within a redshift separation $|\Delta z| < 250$ km s$^{-1}$ are shown. Halos are limited to the population hosting galaxies with M$_* > 10^{11} \Msun$ including all halos (\textit{left}) and centrals only (\textit{right}).}
     \label{fig:Counts2}
\end{figure*}

\subsection{Counts in Cylinders Feature Importance}\label{subsec:CoC_FI}

Figure \ref{fig:Counts2} shows the average halo mass as a function of neighbor counts for halos hosting galaxies with M$_* > 10^{11} \Msun$. The neighbor counts are limited to the $|\Delta z| = 0 - 250$ km s$^{-1}$ bin, which proved most relevant to the network. When considering both centrals and satellites (left), there is some small, non-monotonic trend between average halo mass and neighbor counts, but for every cylinder radius, the change in average halo mass over the full range of neighbor counts is not dramatic ($< 0.4$ dex), especially when compared with the large scatter. When we limited the analysis to central halos (right), the 0.5 $h^{-1}$Mpc cylinder (and to a lesser extent the 1 $h^{-1}$Mpc cylinder) display a more significant evolution ($\sim 1$ dex) in average halo mass with neighbor counts. This suggests the smaller cylinders contain more information about halo mass for central halos.

To analyze the relative importance of different information from the counts in cylinders, we performed a feature removal process. This process follows the same general method as described in Section \ref{subsec:kNN_Features} for the nearest neighbors measure. 
First, we grouped the counts in cylinder data by cylinder radius, masking information from the larger cylinders while retaining the full range of redshift bins.

\begin{figure}[h]
     \centering
     \includegraphics[width=0.45\textwidth]{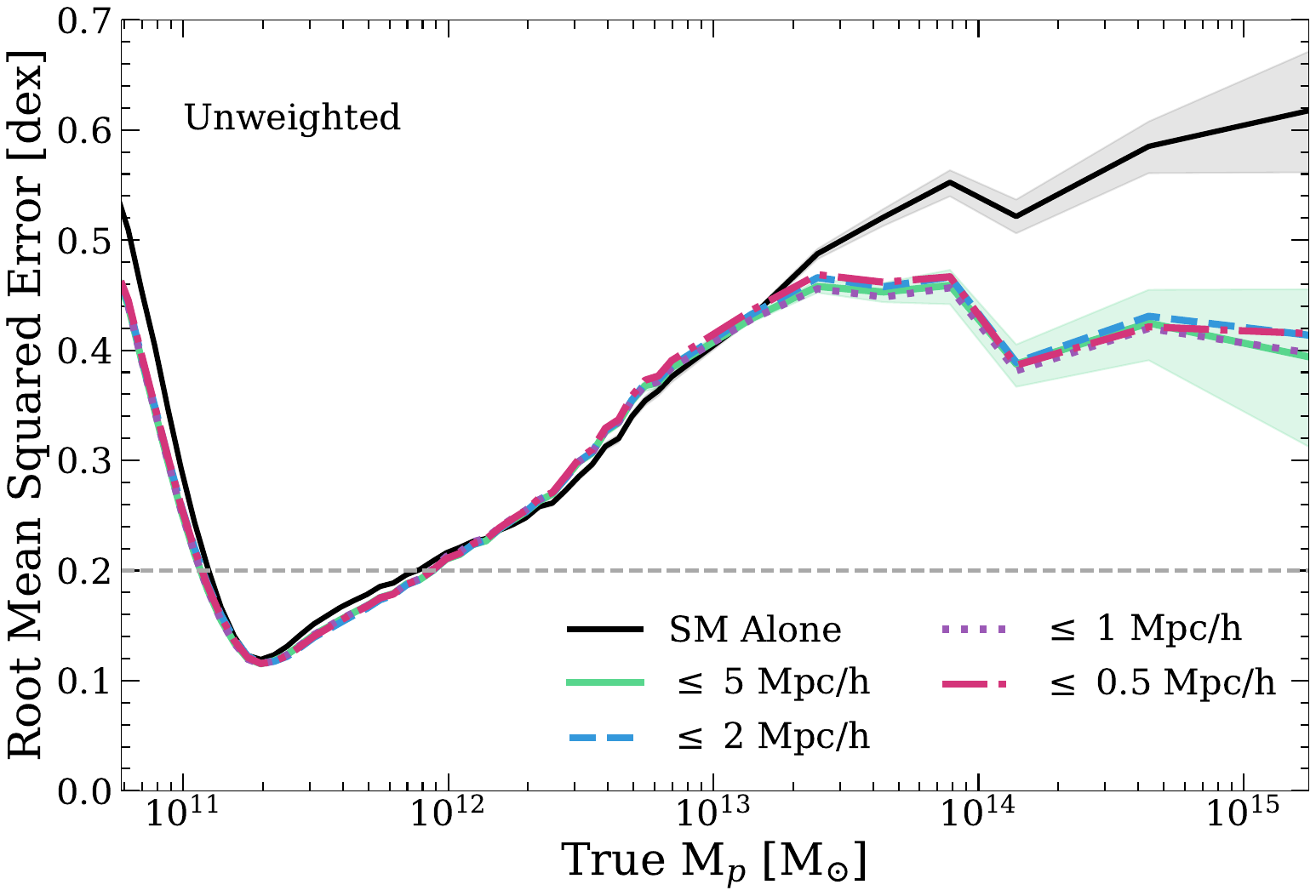}
     \includegraphics[width=0.45\textwidth]{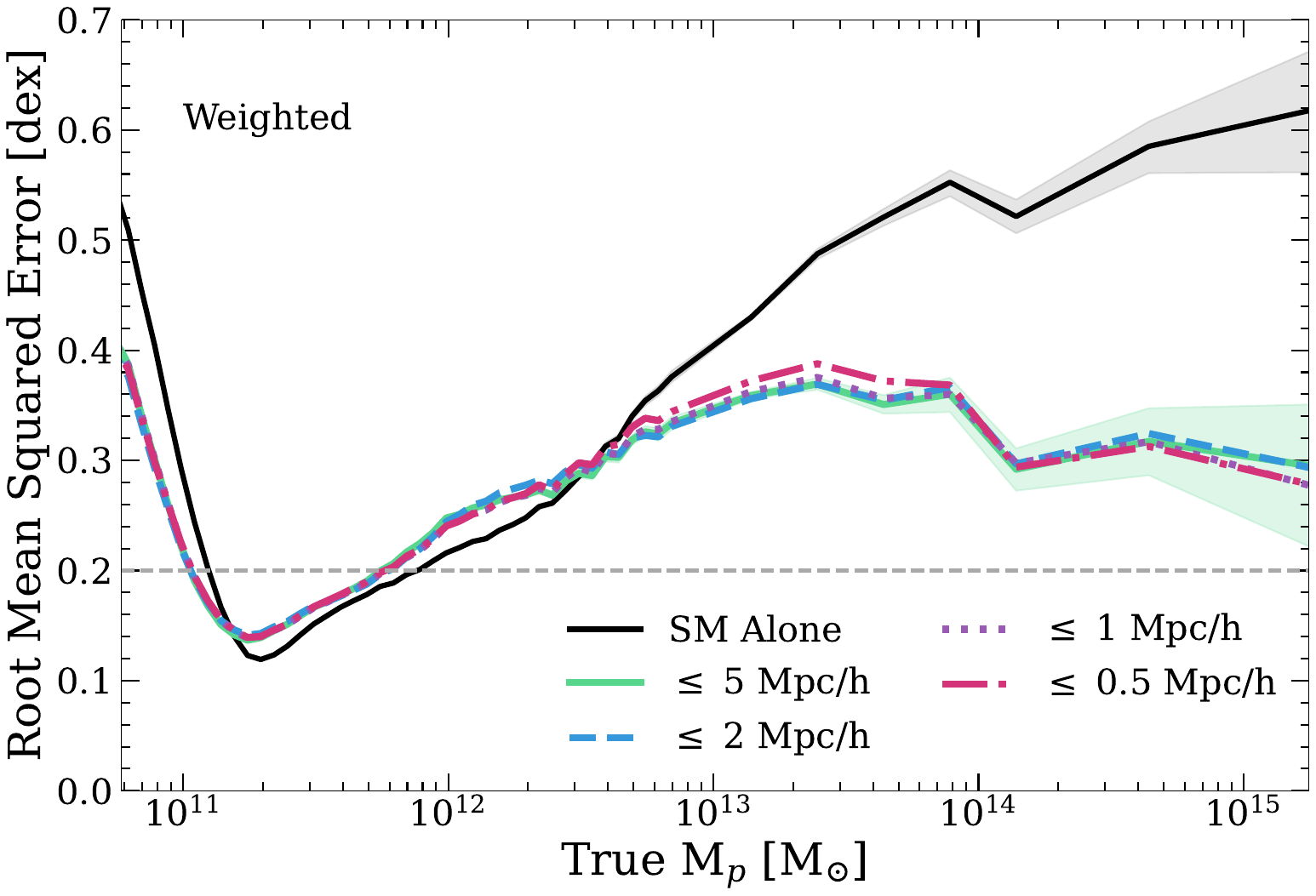}
     \caption{Loss (RMSE in dex) of unweighted (top) and weighted (bottom) networks is plotted as function of halo mass for the four unweighted counts in cylinders networks trained on the full information, with the 5 $h^{-1}$Mpc cylinder masked, with the 2 and 5 $h^{-1}$Mpc cylinders masked, and with the 1, 2, and 5 $h^{-1}$Mpc cylinders masked (green, blue, purple, and pink respectively). The black solid line shows the loss from the SHMR interpolation (i.e., using information from stellar mass alone) for comparison. Shaded regions show 1$\sigma$ errors as estimated from bootstrap resampling. The error regions of the other lines are excluded for clarity but are of similar magnitudes to the ones shown. A horizontal dashed line shows 0.2 dex accuracy for reference.}
     \label{fig:Cyl2}
 \end{figure}

In trial one, we masked the 5 $h^{-1}$Mpc cylinder by replacing the number of counts with zero. In trial two, both the 5 $h^{-1}$Mpc and 2 $h^{-1}$Mpc cylinders were set to zero. Finally, in trial three, every cylinder excluding the 0.5 $h^{-1}$Mpc cylinder was masked. Each iteration is executed with the same network structure and hyperparameters as the full cylinders case. Figure \ref{fig:Cyl2} shows the resulting errors in predictions for the unweighted (top) and weighted (bottom) networks, with the full information shown in green and trials one, two, and three shown by the blue, purple, and pink lines respectively.

 \begin{figure}[h]
     \centering
     \includegraphics[width=0.45\textwidth]{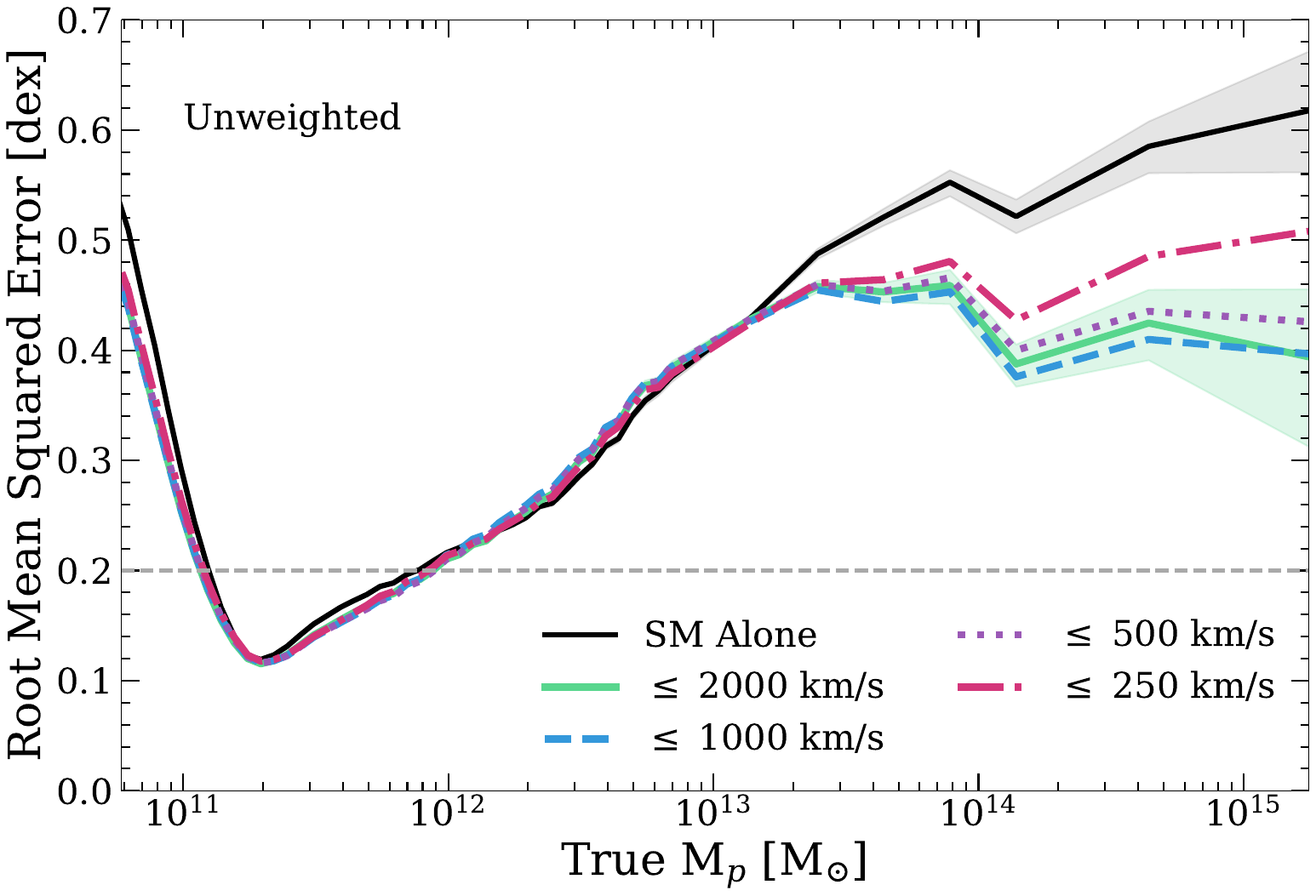}
     \includegraphics[width=0.45\textwidth]{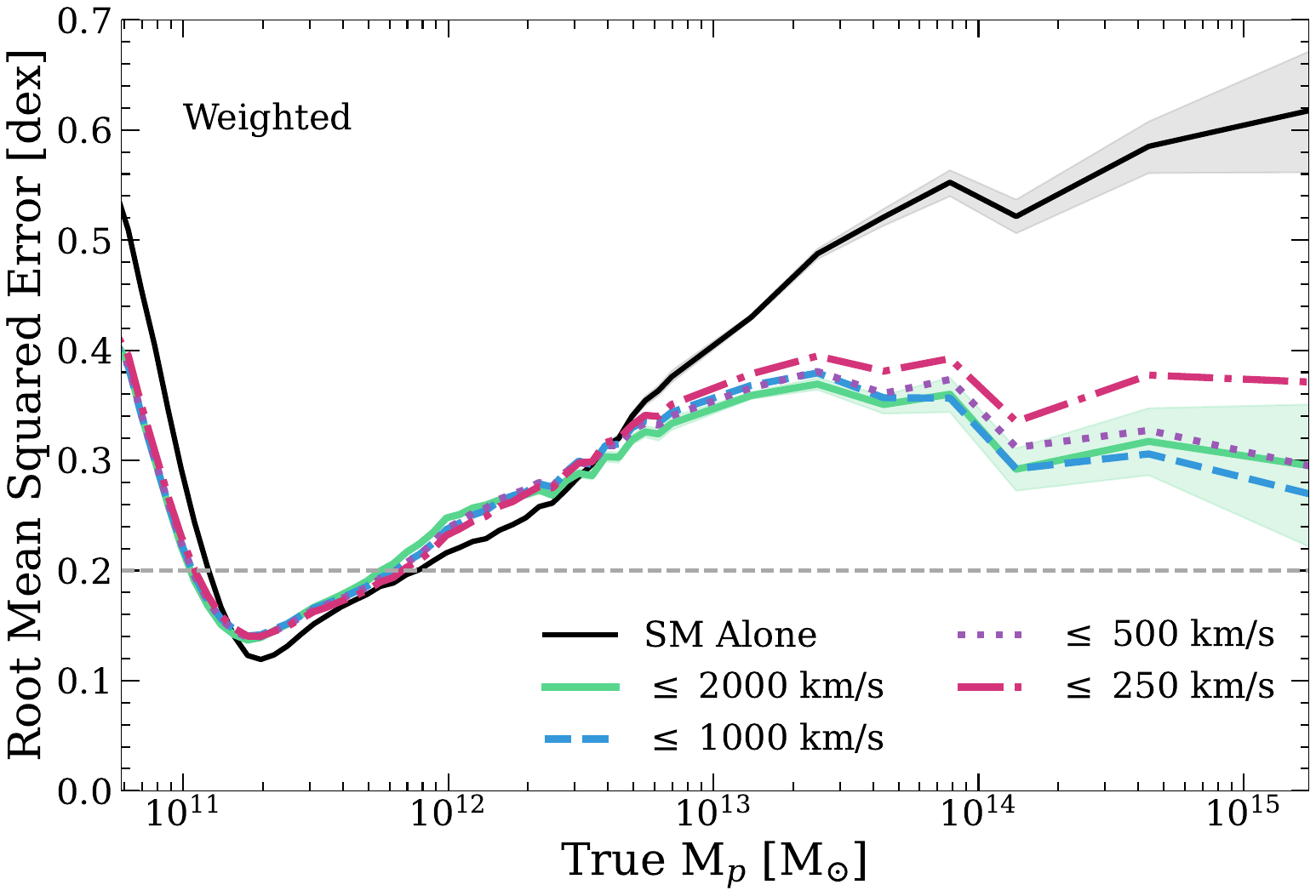}
     \caption{Loss (RMSE in dex) of the unweighted (top) and weighted (bottom) networks is plotted as a function of halo mass for the counts in cylinders networks trained on the full information, with redshift bins $\geq$ 1000 km s$^{-1}$ cylinder masked, $\geq$ 500 km s$^{-1}$ masked, and $\geq$ 250 km s$^{-1}$ masked (green, blue, purple, and pink respectively). The black solid line shows the loss from the SHMR interpolation (i.e., using information from stellar mass alone) for comparison. The shaded region shows 68\% scatter as estimated from bootstrap resampling. The error regions of the other lines are excluded for clarity but are of similar magnitudes to the ones shown. A horizontal dashed line shows 0.2 dex accuracy for reference.}
     \label{fig:Cyl3}
 \end{figure}

In both the unweighted and weighted cases, all four networks have highly similar overall losses. Even with predictions divided into halo mass bins, the networks show no distinguishing characteristics. This suggests that the majority of the information is in the stellar mass and the nearby environment (as represented by the 0.5 $h^{-1}$Mpc cylinder). Beyond this inner region, the network as designed is not extracting any significant information about halo mass.

In addition to considering the projected area covered by the cylinders, we also analyzed the information in each redshift bin, by masking out different bins (Figure \ref{fig:Cyl3}).
In trial one, we masked the bins with redshift separation $|\Delta z| > 1000$ km s$^{-1}$ (blue dashed line) by setting the counts in each bin to zero. In trials two (purple dotted line) and three (pink solid line) we masked out all bins with separation $|\Delta z| > 500$ km s$^{-1}$ and $|\Delta z| > 250$ km s$^{-1}$ respectively. 

Figure \ref{fig:Cyl3} shows results for the unweighted networks (top) and weighted networks (bottom). In both cases, there is no significant change in the loss as a result of removing all redshift bins $|\Delta z| > 500$ km s$^{-1}$. The additional masking of the $|\Delta z| = 250-500$ km s$^{-1}$ in trial three produces a network that is slightly less accurate ($< 0.1$ dex difference) at the high-mass end. These results suggest the majority of the information relevant to the network is contained in the smallest redshift bin, $|\Delta z| = 0 - 250$ km s$^{-1}$, with some potential additional information in the $|\Delta z| = 250-500$ km s$^{-1}$ bin. As in the aperture size tests, the innermost region tested appears to contain the majority of the relevant information on halo mass. Considerations of relative performance on central and satellite halos is reserved to the following section.

\subsection{Combination Network Results}\label{subsec:Combo}

From Figure \ref{fig:Combo1}, we saw that the weighted $k$NN and counts in cylinders networks have similar errors in their halo mass predictions for the same halos. The small amount of scatter ($\lesssim 0.5$ dex) in the prediction errors of the two networks decreases towards larger errors, suggesting that a network combining the information from the two environmental measures will likely not be much more accurate than one using the individual measures. To see whether any additional information can be extracted, we combine the input information from the two environmental measures to create a new network that takes 134 inputs.

\begin{figure}[h]
     \centering
     \includegraphics[width=0.45\textwidth]{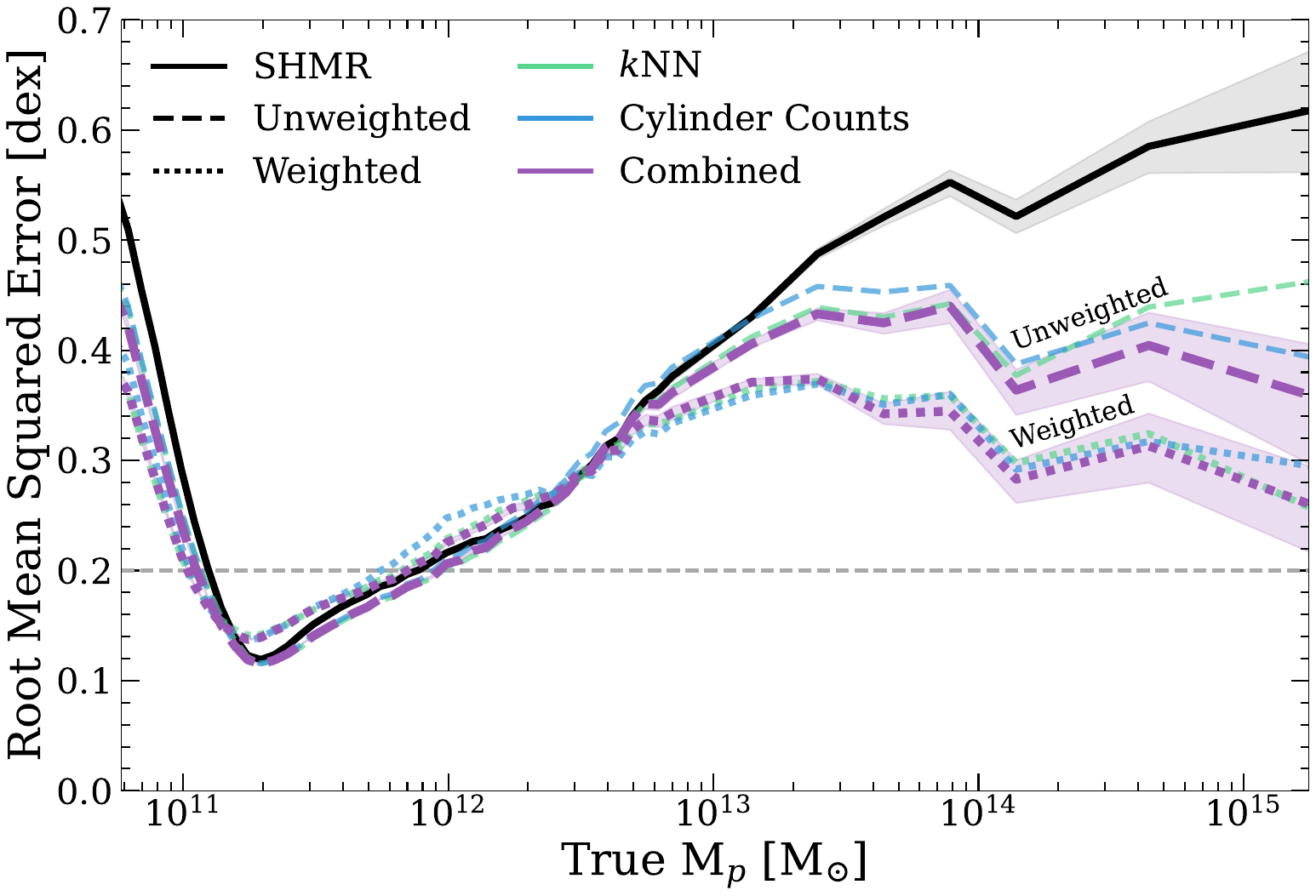}
     \caption{Loss (RMSE in dex) is plotted as a function of halo mass for the unweighted (purple dashed line) and weighted (purple dotted line) combination networks. The performance of the SMHR interpolation from Section \ref{subsec:SM_Alone} is shown in black for comparison. Shaded regions show 1$\sigma$ errors as estimated from bootstrap resampling. A horizontal dashed line shows 0.2 dex accuracy for reference.}
     \label{fig:Combo2}
 \end{figure}

Figure \ref{fig:Combo2} shows the performance of a network trained on the combined information (purple) compared with the individual $k$NN (green) and cylinder counts (blue) networks. There is little change in the performance of the combined network compared to either of the single metric networks. The overall error of the unweighted network is 0.19 dex, which is a $\lesssim 1\%$ difference from the loss of the individual unweighted $k$NN and cylinder counts networks. The weighted network had an overall error of 0.20 dex, which is similarly a negligible change from the performance of the weighted $k$NN and cylinder counts networks. The performance is also similar to the previous models when limited to halo masses $M_p > 10^{13} \Msun$, where it has an RMSE of 0.36 dex. This is a 27\% improvement on the stellar mass alone estimates. 

While only small changes are observed between the individual networks and the combined network, the weighted combined network does provide the lowest overall loss of all networks tested. Hence, we retain the combined network as our primary model for the following analysis and intend to use this as the fiducial model for future work on secondary halo properties.

\begin{figure}[h]
     \centering
     \includegraphics[width=0.45\textwidth]{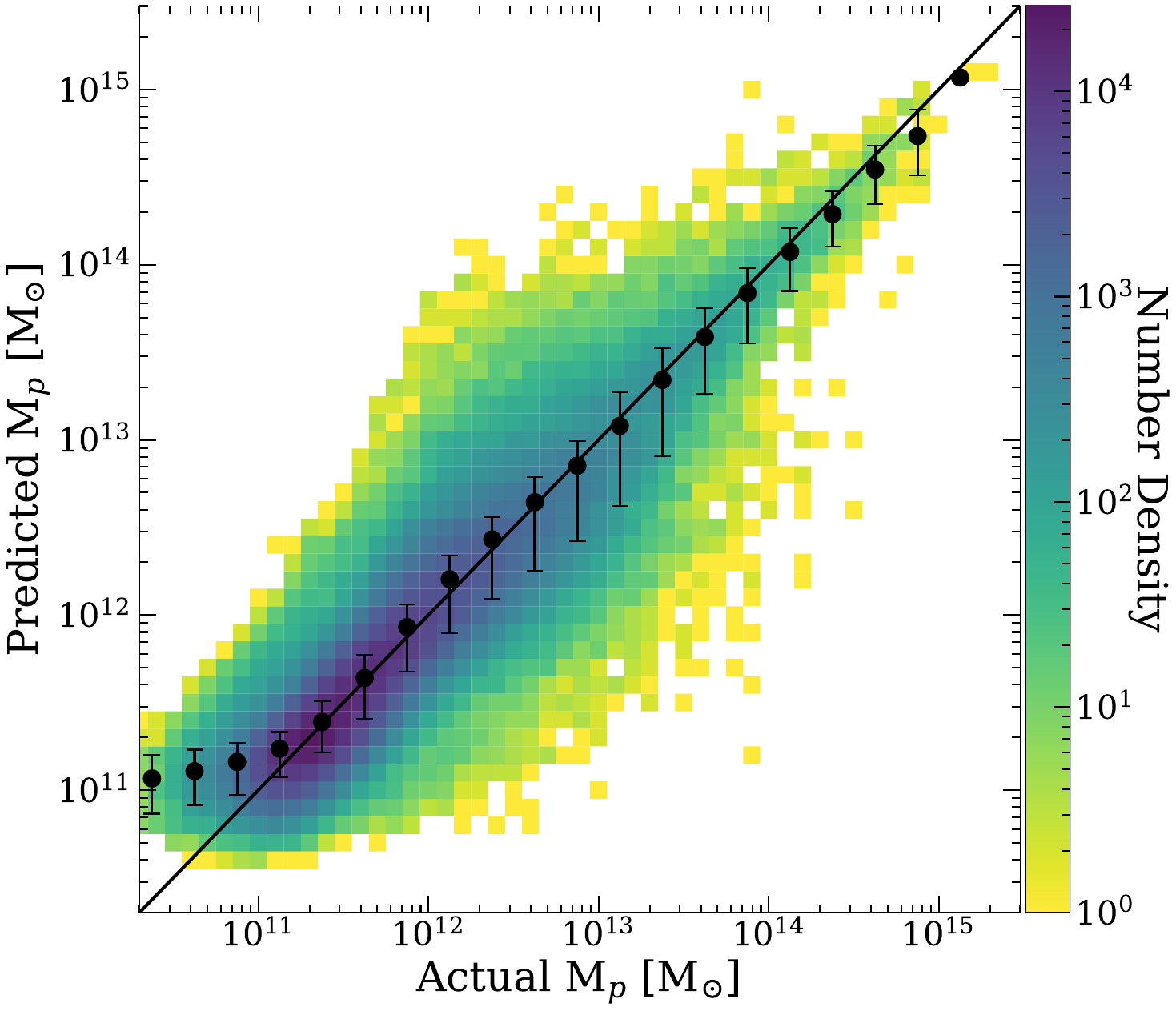}
     \caption{The 2D histogram displays predicted halo masses from the weighted combined network as a function of actual halo mass, with the density of objects represented by the color. Black points show the median predicted halo masses for given actual halo mass bins with one sigma error bars.}
     \label{fig:Combo3}
\end{figure}

The halo masses predicted by the weighted combined network are plotted against the true halo mass values in Figure \ref{fig:Combo3}. The network is highly accurate, with median predictions (as shown by black points) falling mainly on the one-to-one line with a reduced level of scatter compared to the SHMR interpolation (as shown by the size of the error bars). The network tends to overpredict the masses of low-mass halos and underpredict the masses of high-mass halos, similar to the SHMR interpolation. Weighting the training data has reduced this bias, but not fully eliminated it. 

\begin{figure}[h]
     \centering
     \includegraphics[width=0.45\textwidth]{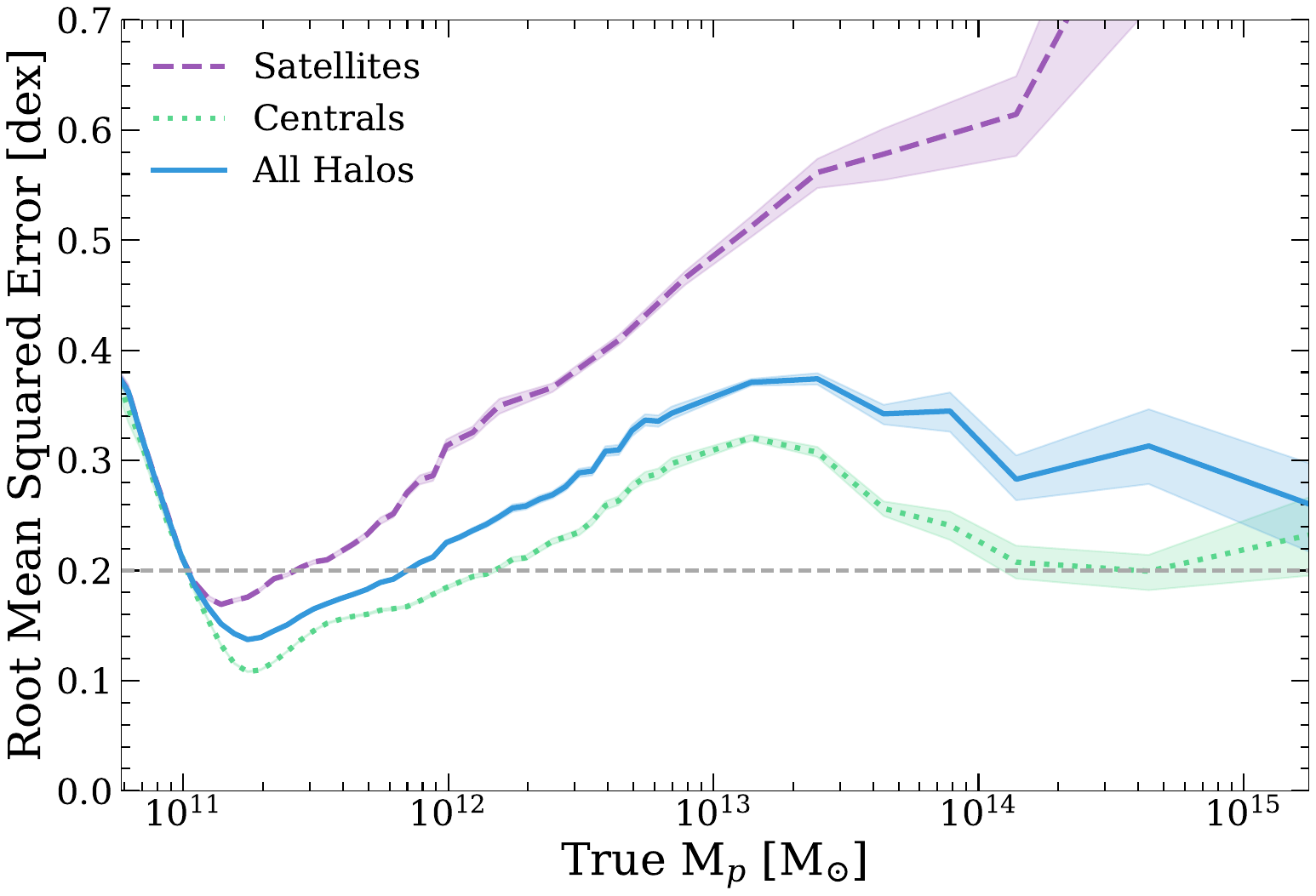}
     \caption{Loss (RMSE in dex) is plotted as a function of true halo mass for the weighted combined network. The population is split into satellites (purple dashed line) and centrals (green dotted line). The performance over the whole population is shown by the blue solid line. Shaded regions show 1$\sigma$ errors as estimated from bootstrap resampling. A horizontal dashed line shows 0.2 dex accuracy for reference.}
     \label{fig:Combo4}
 \end{figure}

We additionally evaluated the performance of the weighted combined network on centrals and satellites separately (Figure \ref{fig:Combo4}). We performed the same separation analysis on the $k$NN and counts in cylinders networks individually. In each case, the results were highly similar to those found for the combination network.
The network is highly accurate for central halos alone, with an overall error of 0.17 dex and an average loss of $\lesssim 0.3$ dex in all mass bins. Excluding the $M_p < 10^{11} \Msun$ regime, the peak in error is at $M_p \sim 10^{13} \Msun$. This is above the turning point in the SHMR, where we observe greater scatter in halo mass at fixed stellar mass, yet below the regime where the satellite information is expected to become highly informative.

On the other hand, for satellite halos, the network loss diverges with increasing halo mass. This is expected as the local environmental information will be more closely tied to the mass of the host halo of the satellite rather than the mass of the satellite halo itself. Thus, there is contrasting information provided to the network by the stellar mass of the satellite galaxy and the neighboring galaxy density. The information contained in the environment would likely be different if we had instead studied the larger composite dark matter halo containing the low-mass subhalo. However, when considering solely the mass of the satellite halo, there is no substantial improvement in the overall loss of the weighted network on satellites compared to the stellar mass-only estimates. 
When taken together with the results in Sections \ref{subsec:kNN_Features} \& \ref{subsec:CoC_FI}, this suggests that even at the field-level of the galaxy distribution, the available information about halo mass is fundamentally limited, and is entirely contained by a small handful of summary statistics covering the immediate environment of the galaxy.

\section{Discussion}\label{sec:Discussion}

The aim of this paper is to address what information regarding halo mass is present in different parts of the halo's observable environment. We find that, beyond the stellar mass of the hosted galaxy, information about the nearest neighbors in the innermost region around a halo's center ($\lesssim 1$ Mpc) is the most informative. This information is similarly contained in measurements expressed through either distance to the halo's nearest neighbors or counts of neighbors in cylinders surrounding the halo.

Our results indicate that at low halo masses ($M_p \lesssim 10^{12.5} \Msun$), the environment contains little supplemental information about the target's host halo mass above and beyond the target's stellar mass. At higher halo masses, the information content of the environment increases, and including data about the distribution of nearby galaxies can improve halo mass estimates (see Section \ref{subsec:kNN}).

The neural networks trained on distances to nearest neighbors and on counts in cylinders had markedly similar performances, as evidenced by the alignment of their prediction errors (Figure \ref{fig:Combo1}). We expected that the nearest neighbors' distances would be the more sensitive of the two probes on small scales, and thus more relevant for halos with $M_p \sim 10^{12.5}-10^{13.5} \Msun$ that likely have few satellites. The 5th nearest neighbor is often used as a probe of the environment in the literature (e.g., \citealt{Muldrew2012, Kawinwanichakij2017, Lacerna2018, Gargiulo2019}). Our results support the choice of the distance to the $k = 5$ neighbor for probing the environment on scales that are sensitive to halo mass. Smaller values of $k$ are more prone to noise, while larger values tend to probe the larger-scale environment that is less sensitive to halo mass. Overall, it appears the two different environmental measures contain the same information, with that information primarily concentrated at small distance scales. 

Additionally, we expected the cylinder counts to be more helpful than the neighbor distances on the more massive end ($M_p > 10^{13.5} \Msun$) where halos have an abundance of satellites. This is evidenced by the relationship between counts in cylinders and halo mass for $M_p > 10^{13.5} \Msun$ within the smaller radii bins (Figure \ref{fig:Cylinders}), while the trend between halo mass and projected distance to the $k$th nearest neighbor is more ambiguous (Figure \ref{fig:Neighbors}). However, in both cases, the high number of low-mass halos found in high-density environments, particularly when including satellite halos, results in significant overlaps between high- and low-mass objects for a given environmental parameter (see Figures \ref{fig:Dists2} and \ref{fig:Cyl2}).

Both environmental measures likely share similar sources of error. For example, an increase in the density of neighboring galaxies as a result of projection effects would result in a similar impact to the two measures. The correlation in prediction loss on individual galaxies (Fig. \ref{fig:Combo1}) supports the idea that error is primarily driven by the same sources across the two network types. 

We can also consider the particular cases where the networks fail to capture the true mass of a halo. For this purpose, we looked for objects with errors in halo mass estimation of greater than 1.0 dex. This corresponds to  0.13\% of the objects in the \textit{Bolshoi-Planck} dataset in the case of the full kNN network, with similar failure rates for the cylinder counts and combination networks. Already, these values indicate a low failure rate for the networks on individual objects. 

With further consideration, the vast majority of these objects ($\sim 92$\%) correspond to halos with SHMR more than five standard deviations from the average relationship. This includes, for example, a $\sim 10^{14}\Msun$ halo hosting a galaxy with a stellar mass of $\sim 10^{10}\Msun$. We suspect the mass values assigned to many of these halos are the result of a bug in \textsc{UniverseMachine} or in the merger tree construction. In order to avoid the suspected anomalous halos, we considered removing objects that fell more than five standard deviations from the average SHMR. However, training and testing on datasets sampled in this manner resulted in no significant changes in network performance overall. Hence, the SHMR outliers were retained in all other calculations.

This study differs from previous ML-based halo mass estimates due to the inclusion of satellite halos in the target population. We found drastically different performances between centrals and satellites. As local galaxy density is expected to scale with the mass of the central halo, it is less informative about the masses of the satellites. Thus, it is not unexpected that our best networks, which perform well on centrals ($\lesssim 0.3$ dex mean squared error), do not capture the halo mass of satellites to the same level of accuracy. While past studies have focused on the masses of central halos (e.g., \citealt{Ntampaka2015, Ho_2019, Calderon2019}), we considered both satellites and centrals together due to the difficulty of fully separating the two populations in observations. Appendix \ref{sec:AppA} demonstrates preliminary results for predicting the halo mass of centrals assuming a perfect separation of central and satellite halos. Given the substantial difference in satellite and central behavior, separating the two categories based on observational data will likely be important for future work probing the mass and secondary properties of halos. An ML study to find the most accurate method for separating centrals and satellites would be an important milestone for understanding halo properties in observations. 

The neural networks presented in this paper are designed to be applied to observational surveys, but several factors should be considered prior to such an application. Our training and testing are conducted on fully complete and clean data. Real observational surveys will include fiber collisions and other sources of error or incompleteness to which the network may or may not be robust. Hence, it is important to apply the networks to a relatively clean data set. However, in the realistic limit that no data set is perfectly clean, one should test the robustness of the network against perturbations on the scale of the error expected in the observed data. In addition, it is worth noting that it may be simpler to correct cylinder counts data for fiber incompleteness than the distance to the $k$th nearest neighbor. Therefore, the cylinder counts network may be more easily applied to an observational survey. 

Future papers in this series are planned to address secondary halo properties such as concentration, mass accretion history, and time since the last major merger. The findings of this paper reemphasized the importance of separating centrals and satellites when investigating trends between secondary halo properties and the environment, as the environment may correlate less with satellite properties. In addition, our results highlight the relationship between the local environment and halo mass, which will need to be marginalized over when considering secondary halo properties.

\section{Conclusions}\label{sec:Conclusions}

Our main conclusions are summarized as follows:
\begin{enumerate}
    \item Stellar mass alone is a strong predictor of halo mass, containing far more information than is found in the local galaxy distribution. This is especially clear for $M_p \lesssim 10^{12.5} \Msun$, where there is no clear improvement gained from including information about the environment. Above this mass threshold, the inclusion of environmental information becomes more significant with increasing halo mass (Sections \ref{subsec:kNN} and \ref{subsec:CoC_Results}).
    \item Information about halo mass is extremely spatially restricted, with the innermost regions ($\sim$ 1 Mpc) surrounding the center of the target halo containing the majority of the information about halo mass. This is demonstrated by the similar performances of both the $k$NN and cylinder counts networks with larger-scale information masked out in comparison with the performances of the full networks (Sections \ref{subsec:kNN_Features} and \ref{subsec:CoC_FI}).  
    \item The performance of the $k$NN and cylinder counts networks are remarkably similar, suggesting that both environmental measures contain the same information about halo mass (Section \ref{subsec:CoC_Results}).
    \item The weighted combination network was the best performing model, with a slightly lower error than the $k$NN and cylinder counts networks alone. We plan to use this as the fiducial model for future inquiries into halo properties.
\end{enumerate}

\section*{Acknowledgements}

We thank Tom Abel, Han Aung, Aleksandra Ciprijanovic, Tim Eifler, Xiaohui Fan, Robert Feldman, ChangHoon Hahn, Christian Kragh Jespersen, Chris Lovell, Josh Peek, Joel Primack, Risa Wechsler, John Wu, Ann Zabludoff, and Haowen Zhang for insightful discussion during the development of this paper. 

HB and PB were funded through a Fellowship from the Packard Foundation, Grant \#2019-69646. Work done by APH at Argonne National Laboratory was supported under the DOE contract DE-AC02-06CH11357. This research was supported in part by the National Science Foundation under Grant No. NSF PHY-1748958. This research is based upon High Performance Computing (HPC) resources supported by the University of Arizona TRIF, UITS, and Research, Innovation, and Impact (RII) and maintained by the UArizona Research Technologies department. The University of Arizona sits on the original homelands of Indigenous Peoples (including the Tohono O’odham and the Pascua Yaqui) who have stewarded the Land since time immemorial.

The Bolshoi-Planck simulation was performed by Anatoly Klypin within the Bolshoi project of the University of California High-Performance AstroComputing Center (UC-HiPACC; PI Joel Primack). Resources supporting this work were provided by the NASA High-End Computing (HEC) Program through the NASA Advanced Supercomputing (NAS) Division at Ames Research Center.

The SMDPL simulation was performed by Gustavo Yepes on the SuperMUC supercomputer at LRZ (Leibniz-Rechenzentrum) using time granted by PRACE, project number 012060963 (PI Stefan Gottloeber).

\section*{Data Availability}

Trained models, as well as the codes used to create them, are available online at \url{https://github.com/hbowden-arch/HaloProperties}.
\textsc{UniverseMachine} galaxy catalogs \citep{UM} can be found at \url{https://www.peterbehroozi.com/data.html}.

\bibliographystyle{mnras}
\bibliography{sources}


\appendix

\section{Centrals only training}\label{sec:AppA}

For ease of comparison with other works, we include here the results of networks trained and tested on only central halos. These results assume zero contamination from satellite halos in the test population. The \textit{Bolshoi-Planck} simulated test set can be perfectly separated into central and satellite populations, but this level of accuracy is not available for observational surveys. 

\begin{figure}[h]
     \centering
     \includegraphics[width=0.45\textwidth]{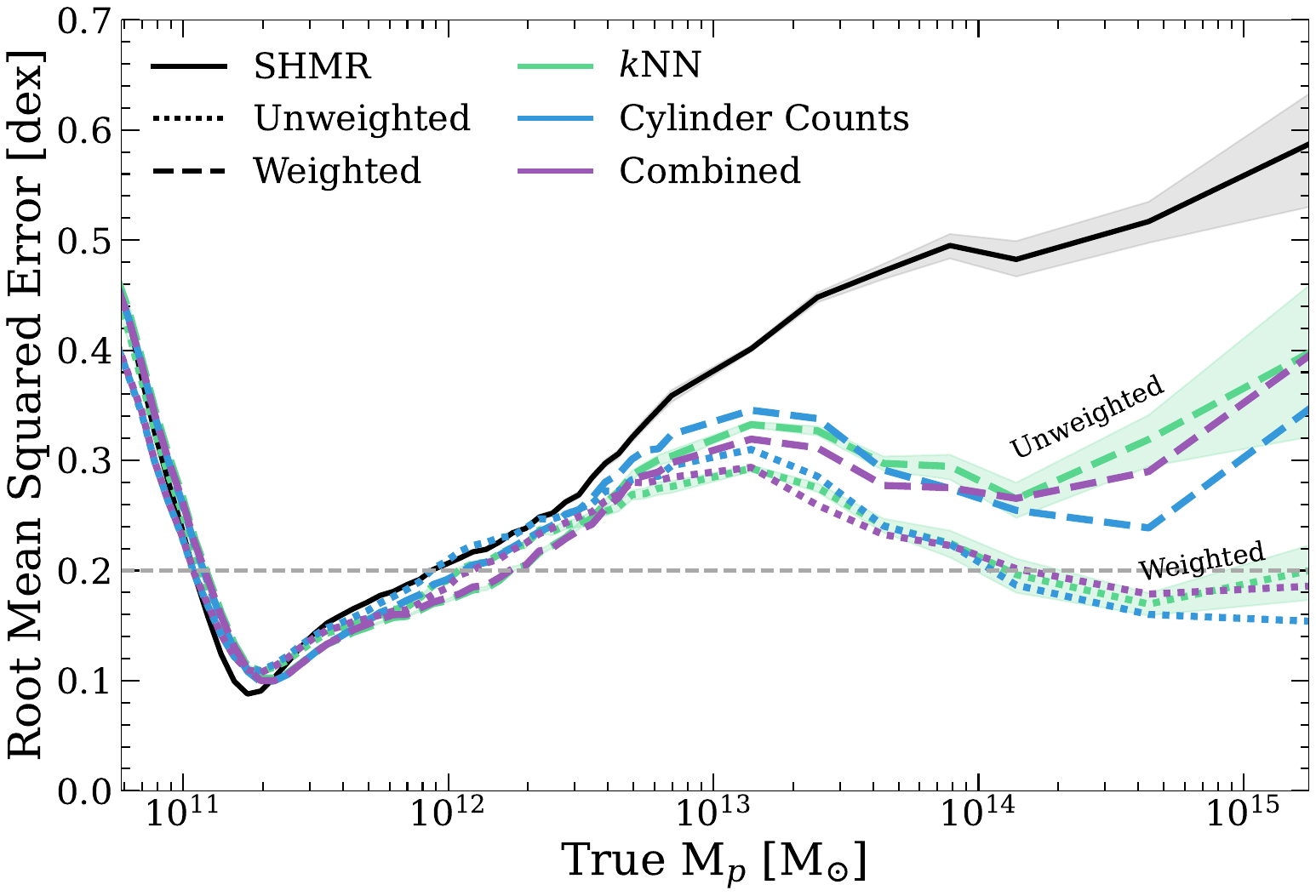}
     \caption{For networks trained and applied to only central halos, loss (RMSE in dex) is plotted as a function of halo mass for unweighted (solid) and weighted (dashed) networks trained on 50 nearest neighbors (green), cylinder counts (blue), and a network with combined inputs (purple). The performance of the SMHR interpolation is shown in black for comparison. The shaded black and green regions signify the 1$\sigma$ errors estimated from bootstrap resampling corresponding to the SMHR interpolation and the $k$NN network, respectively.  The error regions of the other lines are excluded for clarity but are of similar magnitudes to the ones shown. A horizontal dashed line shows 0.2 dex accuracy for reference.}
     \label{fig:App1}
 \end{figure}

Figure \ref{fig:App1} shows the resulting prediction errors in peak halo mass for the SHMR interpolation method as well as the weighted and unweighted nearest neighbors and cylinder counts networks trained and tested on central halos. All the tested neural networks outperform the SHMR for halos with $M_p \gtrsim 10^{12} \Msun$. The overall RMSE for all networks is $\sim 0.16-0.17$ dex. As found in the case with all halos included, the weighted networks (dotted lines) outperform the unweighted networks (dashed lines) for $M_p \gtrsim 10^{13} \Msun$. No substantial difference between the performance of the different types of inputs is evident. There is no significant change (< 1\%) in the overall RMSE performance on centrals between the centrals-only networks and the networks trained on centrals and satellites (Figure \ref{fig:App1}). However, when limited to halos with $M_p \sim 10^{14}\Msun$, the centrals-only networks perform $\sim 0.02-0.06$ dex better than the networks trained on a combination of centrals and satellites. It appears that removing the satellites from the training sample had little effect on how the trained network performed on central halos at low halo masses but did improve performance slightly for the most massive halos.

At the cluster mass end (M$_p \sim 10^{14}\Msun$), we can compare the results of our centrals-only neural network approach against a variety of cluster mass recovery techniques such as those evaluated in \cite{Old2015}. They consider the accuracy of over twenty non-ML methods applied to clusters from two galaxy catalogs. The different techniques have a range of root-mean-squared accuracies from $\sim 0.2-0.6$ dex when applied to cluster populations with an average true halo mass of $\sim 10^{14}\Msun$. Selecting halos from our test catalog with $M_p > 10^{13.5}\Msun$ provides a population with a similar average true halo mass. The average RMSE of all three weighted networks on this population is $\sim 0.2$ dex, corresponding to the best performances seen in the \cite{Old2015} review.

\pagebreak

\section{Redshift Information}\label{sec:AppB}

As designed, the $k$NN network includes no information about the redshift separation of the object and its neighbors. This is in contrast to the counts in cylinders network where neighbors are binned by redshift separation. To determine the impact of this discrepancy between the two methods, we consider how the inclusion of redshift information impacts the results of the $k$NN network. To do this, we created an additional network that took as input the stellar masses, projected distances from the object, and redshift separations from the object for each neighbor. This encompasses all the information the $k$NN network takes with the addition of redshift. 

\begin{figure}[h]
     \centering
     \includegraphics[width=0.45\textwidth]{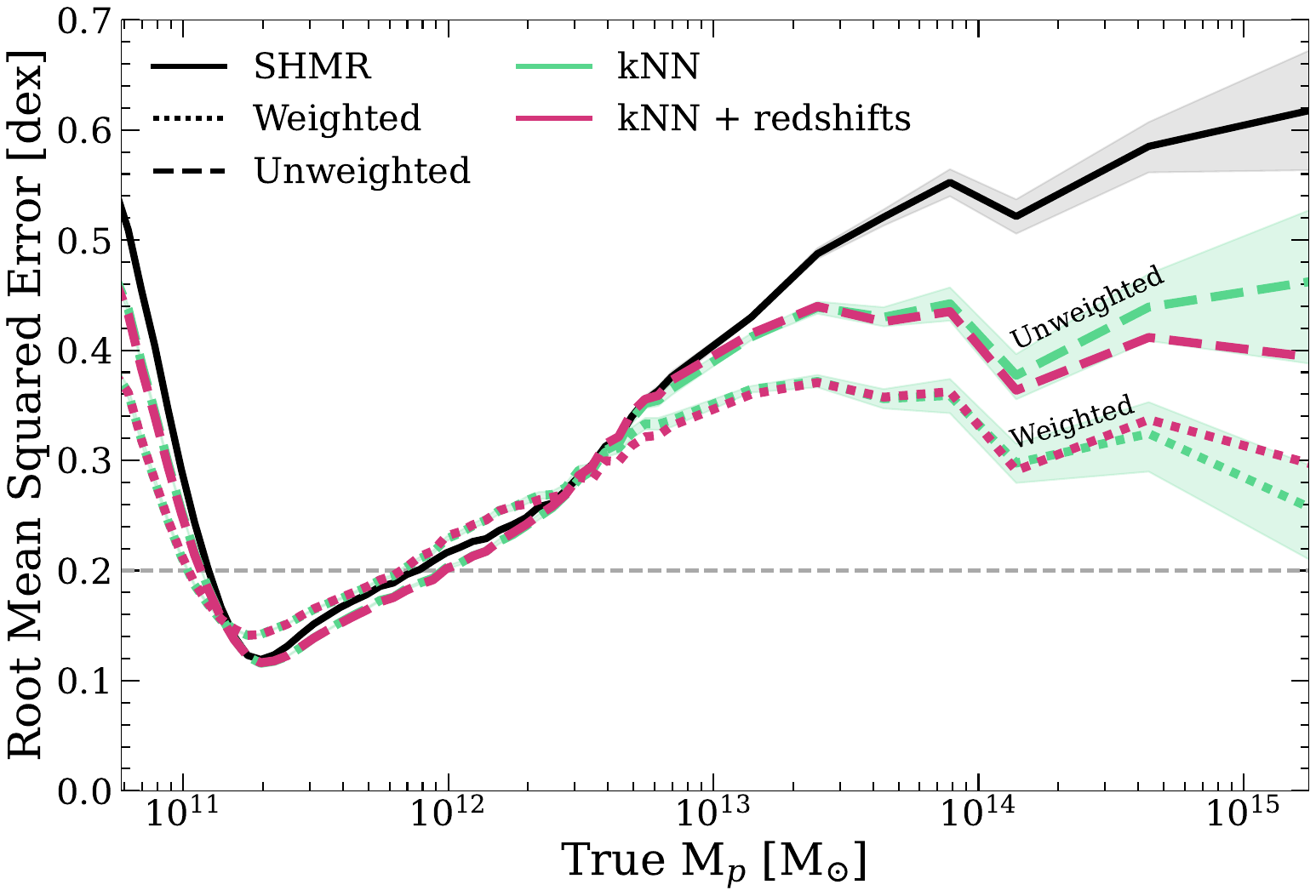}
     \caption{Loss (RMSE in dex) is plotted as a function of halo mass for the unweighted (green dashed line) and weighted (green dotted line) $k$NN networks. The pink lines are added to indicate the results of a network provided with the redshift separation between the object and each neighbor in addition to the full $k$NN information (stellar mass and projected distance from the object). The performance of the SMHR interpolation from Section \ref{subsec:SM_Alone} is shown by the black solid line for comparison. Shaded regions show 1$\sigma$ errors as estimated from bootstrap resampling. A horizontal dashed line shows 0.2 dex accuracy for reference. The performance of the network with redshifts falls within the error regions of the original $k$NN network.}
     \label{fig:App2}
 \end{figure}

Figure \ref{fig:App2} shows the results of the $k$NN network with redshifts (pink) compared to the original $k$NN network (green). The overall accuracies of the weighted and unweighted networks with redshift were comparable to the their counterparts without redshift information. In addition, when considering error as a function of true halo mass, the networks with redshift information fall within the error regions of the original $k$NN networks. Given the lack of clear improvements in network performance when provided with redshift information, we proceeded with the less complex no-redshifts form for the $k$NN models.

\end{document}